\documentclass[11pt]{article}
\usepackage[title,toc,titletoc]{appendix}
\usepackage{titlesec,physics}
\usepackage{float,placeins,subcaption}
\usepackage{amsmath,amssymb,latexsym,lscape,amsthm}
\usepackage[table,xcdraw]{xcolor}
\usepackage[final]{graphicx}
\usepackage[colorlinks=true,linkcolor=black,citecolor=black,urlcolor=blue]{hyperref}
\usepackage[a4paper,top=3cm,bottom=2.5cm,left=2.5cm,right=2.5cm,marginparwidth=1.75cm]{geometry}
\usepackage{multirow}
\usepackage{adjustbox}
\DeclareUnicodeCharacter{0301}{\'{O}}
\usepackage[utf8]{inputenc}
\usepackage{csquotes}
\usepackage[backend=bibtex,style=numeric-comp,sorting=none, maxnames=5]{biblatex}
\definecolor{argentina}{rgb}{0.58, 0.40, 0.74}
\definecolor{bolivia}{rgb}{1.0, 0.84, 0.0}
\definecolor{brazil}{rgb}{1.00, 0.50, 0.05}
\definecolor{repdominican}{rgb}{0.84, 0.15, 0.16}
\definecolor{ecuador}{rgb}{0.17, 0.63, 0.17}
\definecolor{mexico}{rgb}{0.12, 0.47, 0.71}
\definecolor{nepal}{rgb}{0.74, 0.74, 0.13}
\definecolor{thailand}{rgb}{0.89, 0.47, 0.76}

\begin{document}

\begin{center}  {\Large \bf   A Temporal Playbook for Multiple Wave Dengue Pandemic \\~\\ from\\~\\ Latin America and Asia to Italy }
\end{center}
\smallskip
\begin{center}
	Alessandra D'Alise$^{1}$, Davide Iacobacci$^{1}$, Francesco Sannino$^{1,2,3}$ 
\end{center}

\begin{center} 
  {\small \sl $^1$Department of Physics, University of Naples Federico II, INFN Section Naples,\\ via Cintia, I-80126 Naples, Italy }\\
{\small \sl $^2$Quantum  Theory Center ($\hbar$QTC) at IMADA and D-IAS, Southern Denmark University,\\ Campusvej 55, 5230 Odense M, Denmark}\\
{\small \sl $^3$Scuola Superiore Meridionale, Largo S. Marcellino, 10, 80138 Napoli NA, Italy}\\ 
{\small alessandra.dalise@unina.it, davide.iacobacci@unina.it, sannino@qtc.sdu.dk}
\end{center}

\begin{abstract}
We show that the eRG framework is a useful and minimal tool to effectively describe the temporal evolution of the Dengue multi-wave pandemics. We test the framework on the Dengue history of several countries located in both Latin America and Asia. We also observe a strong correlation between the total number of infected individuals and the changes in the local temperature. Our results further support the expectation that global warming is bound to increase the cases of Dengue worldwide.  We then move to investigate, via the eRG, the recent outbreak in Fano, Italy and offer our projections.

\end{abstract}


\newpage

\tableofcontents

%\newpage
%%%%%%%%%%%%%%%%%%%%%%%%%%%%%%%%%%%%%%%%%%%%%%%%%%%%
\section{Introduction}

Pandemics constitute one of the most dangerous health hazards for humans. They can come in different forms depending on the underlying biological mechanisms by which they can affect humans and spread globally. They are affected by a number of factors ranging from thinning the geographical distance across different species to the impact of global warming. Among these, the Dengue fever poses a crucial public health concern and imposes a notable socio-economic burden on numerous tropical and subtropical areas of the world. Transmission to humans occurs through mosquito bites carrying the virus, mainly through the bites of infected \textit{Aedes aegypti} and \textit{Aedes albopictus} mosquitoes \cite{sakamoto2023mathematical}. Among these, \textit{Aedes aegypti} is the most significant vector, known for its high vectorial capacity and strong preference for human blood. This species has adapted well to urban environments \cite{morrison2008defining}, breeding in and around houses in containers and disposed water-holding vessels. As a result, \textit{Aedes aegypti} has extensively spread across the tropics and subtropics, posing challenges for vector control due to their daytime activity and limited flight range. Four serotypes of the Dengue virus \cite{murray2013epidemiology}, belonging to the Flaviviridae family and the genus Flavivirus \cite{gubler2002global, world2011comprehensive}, are responsible for the disease. Infection with any one of these serotypes provides lifelong immunity to that particular serotype \cite{halstead1974etiologies, wilder2010update}. Each serotype has been independently identified as a causative agent of Dengue outbreaks and is associated with more severe Dengue forms. Partially due to global warming, Dengue outbreaks are spreading in the Mediterranean regions such as the example of Fano, in Italy, clearly indicates. 

\begin{figure}[h!]
    \centering
    \includegraphics[scale=0.53]{ 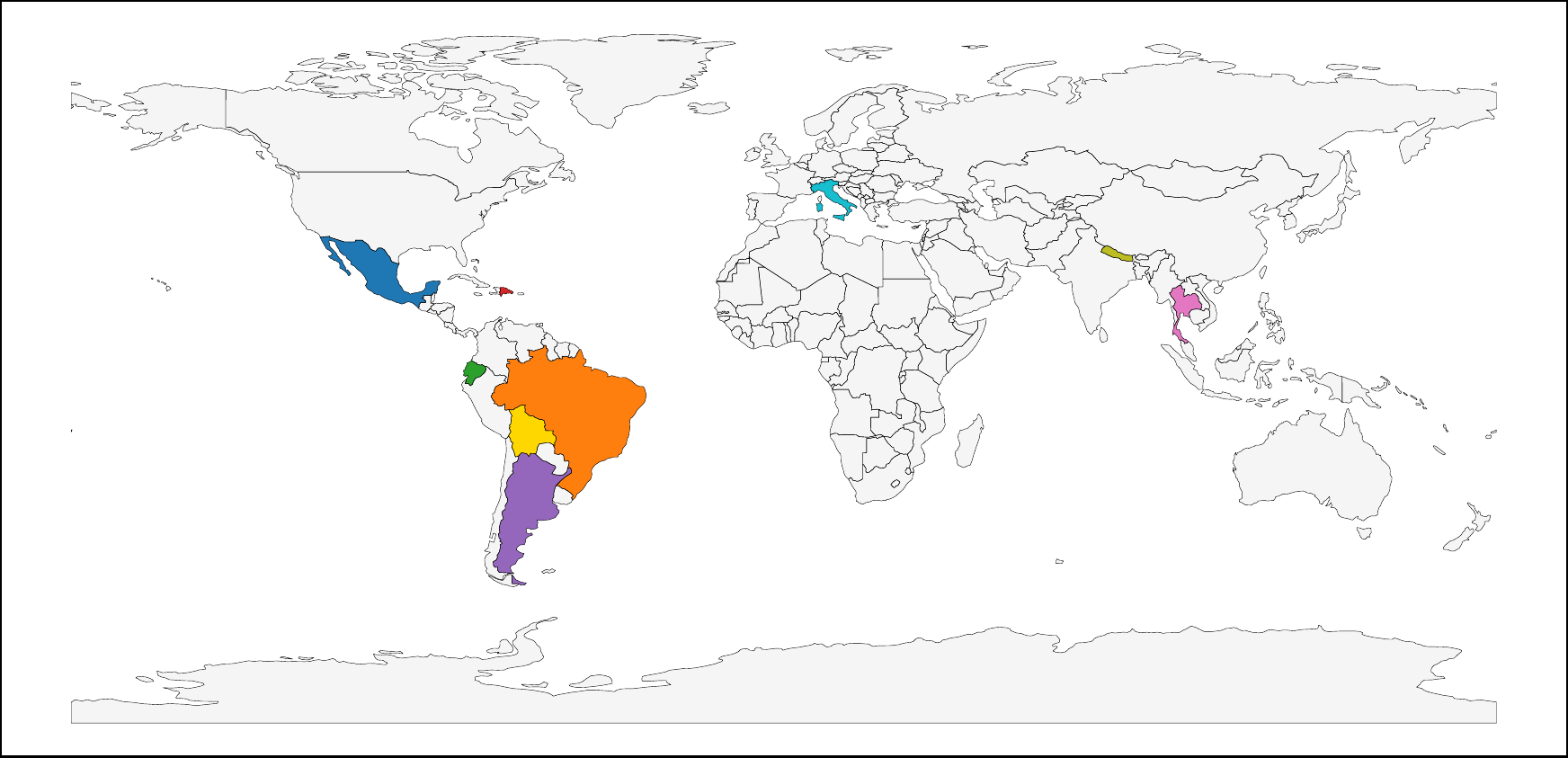}
    \caption{World map illustrating the countries selected for the study of Dengue pandemics. Each country is identified with a specific color: Argentina (purple), Bolivia (yellow), Brazil (orange), Dominican Republic (red), Ecuador (green), Italy (cyan), Mexico (blue), Nepal (olive) and Thailand (pink).}
    \label{fig:GLOBAL}
\end{figure}

Therefore, it is important to develop better tools in support of public health decision makers, as has been done for recent airborne pandemics such as the 2009 H1N1 flu and COVID-19  \cite{coburn2009,Covidreview,james21}.  In fact,  a number of mathematical and computational tools have been developed to analyze data to control and reduce the socio-economic and health impact of airborne pandemics \cite{duan2015,vespignani15,MRC,poletti11}. Among these tools, compartmental models have played an important role. They have matured into quite involved frameworks that often require several parameters, which are sometimes difficult to obtain from observations \cite{merler16,tang2020,trentini2022,SHARMA2021,RODA2020}. Therefore, it is helpful to develop effective models that, with a few parameters, encode the relevant information. The  \emph{epidemiological Renormalization Group} (eRG) framework is one of these example put forward during the COVID-19 pandemics \cite{DellaMorte:2020wlc}.   It was shown to be an effective way to describe the temporal evolution of epidemic waves by organizing compartmental models as RG flows  in between fixed points  \cite{DellaMorte:2020wlc}. Furthermore, in \cite{DellaMorte:2020qry} the link between the eRG and SIR models with time-dependent coefficients was provided. The model has been extended to describe human interaction in different regions of the world \cite{cacciapaglia2020} and used for the well-known prediction of the evolution of the second wave pandemic of COVID-19 in Europe. The eRG framework has been further improved to describe vaccination campaigns in the US \cite{cot2021impact}. A recent development \cite{buonomo2024informationindexaugmentederg} has been the extension of the model to take into account human vaccination behaviors according to the balance between the perceived risk of contagion and the perceived risk of vaccine side effects \cite{TROIANO2021,Lindholt, }. \\
 \\ 
In this work we show that the eRG framework can be readily applied to describe the multiple wave Dengue pandemic globally from Latin American countries to Asian ones. We show in Figure~\ref{fig:GLOBAL} the countries analyzed in this work. We then show that there is a temporal correlation between the variation of the total number of infected individuals and the local temperature once different pandemic waves are analyzed and for different countries. As an European example, we investigate the Dengue outbreak in Fano, Italy. Here we first utilize the eRG to describe the current Dengue wave and use it as a starting point to forecast future waves, taking into consideration the dependence on the local temperature change.

Our paper is structured as follows. In Section \ref{sec:model} we introduce the eRG methdology and summarize its salient points; Section \ref{sec:Latinasiancountries} is dedicated to the study of the multiple Dengue outbreaks in Argentina, Bolivia, Brazil, Ecuador, Dominican Republic and Mexico for the Latin countries and Thailand and Nepal for the Asian ones. At the end of the section we observe how the data on total number of infected are correlated to local changes in temperature. The present and future of Fano Dengue outbreaks is presented in Section \ref{sec:fano}. We offer our outlook and conclusions in Section \ref{sec:conclusions}. In the appendix, we report the result for the eRG fits for the countries investigated in the main text for various pandemic waves. 

\FloatBarrier

\FloatBarrier

\section{Methodology}
\label{sec:model}

 Compartmental models can be derived as macroscopic analogues of microscopic ones \cite{JLCardy_1985}, such as lattice, percolation, and random walks models \cite{Hohenegger:2024xtv}. Among the compartmental models, the eRG framework constitutes an effective approach apt to reproduce the dynamics of epidemic waves while strongly reducing the number of unknown parameters compared to the earlier compartmental models \cite{DellaMorte:2020wlc}. 

Following the renormalization group approach, widely employed in theoretical physics,  the beta function $\beta$ governs the RG time evolution of the interaction strength among the basic constituents of the physical system \cite{DellaMorte:2020wlc}. In epidemiology, $\beta$ is interpreted as the time variation of the epidemic strength $\alpha(t)$, where $\alpha(t)$ is a measure of the averaged interactions among individuals \cite{DellaMorte:2020wlc}. We have that $\beta$ and $\alpha$ are given by:
\begin{equation}
- \beta (\alpha) = \dfrac{d \alpha}{dt}.
\end{equation}
The quantity $\alpha$ is typically assumed  to be the slowly varying function:
\begin{equation}\label{alphaln}
	\alpha(t)=\ln \mathcal{I}(t),
\end{equation}
where $ \mathcal{I}(t)$ is the cumulative number of infectious individuals\footnote{Other monotonic functions of $ \mathcal{I}(t)$ can also  be used \cite{caccia2022}.}. The right-hand side of the $\beta$ equation stems from the expected behavior for a generic epidemic. Typically is a polynomial function of $\alpha$ featuring real and/or complex fixed points \cite{DellaMorte:2020wlc,caccia2022}.

An effective way to determine the right-hand side of the $\beta$ function is to invoke the (approximate) temporal symmetries of the problem. The following equation
\begin{equation}
- \beta (\alpha) = \gamma\alpha  \left( 1 - \dfrac{\alpha}{a} \right),\label{eq:beta0}
\end{equation}
for example, captures the initial and final temporal scale invariance of a single epidemic outbreak.  A single wave may be seen as a flow between two fixed points that represent the beginning of the epidemic, where $\alpha=0$ (say $\mathcal{I}=1$ if (\ref{alphaln}) holds) and the end of the epidemic, when the final size of the epidemic strength is reached, that is $\alpha=a$, respectively. 

\section{Multi-wave dynamics versus global warming for Latin American and Asian Countries}
\label{sec:Latinasiancountries}
\subsection{Latin America}
\begin{figure}[h!]
      \centering
	     \begin{subfigure}{0.45\linewidth}
		 \includegraphics[width=\linewidth]{ 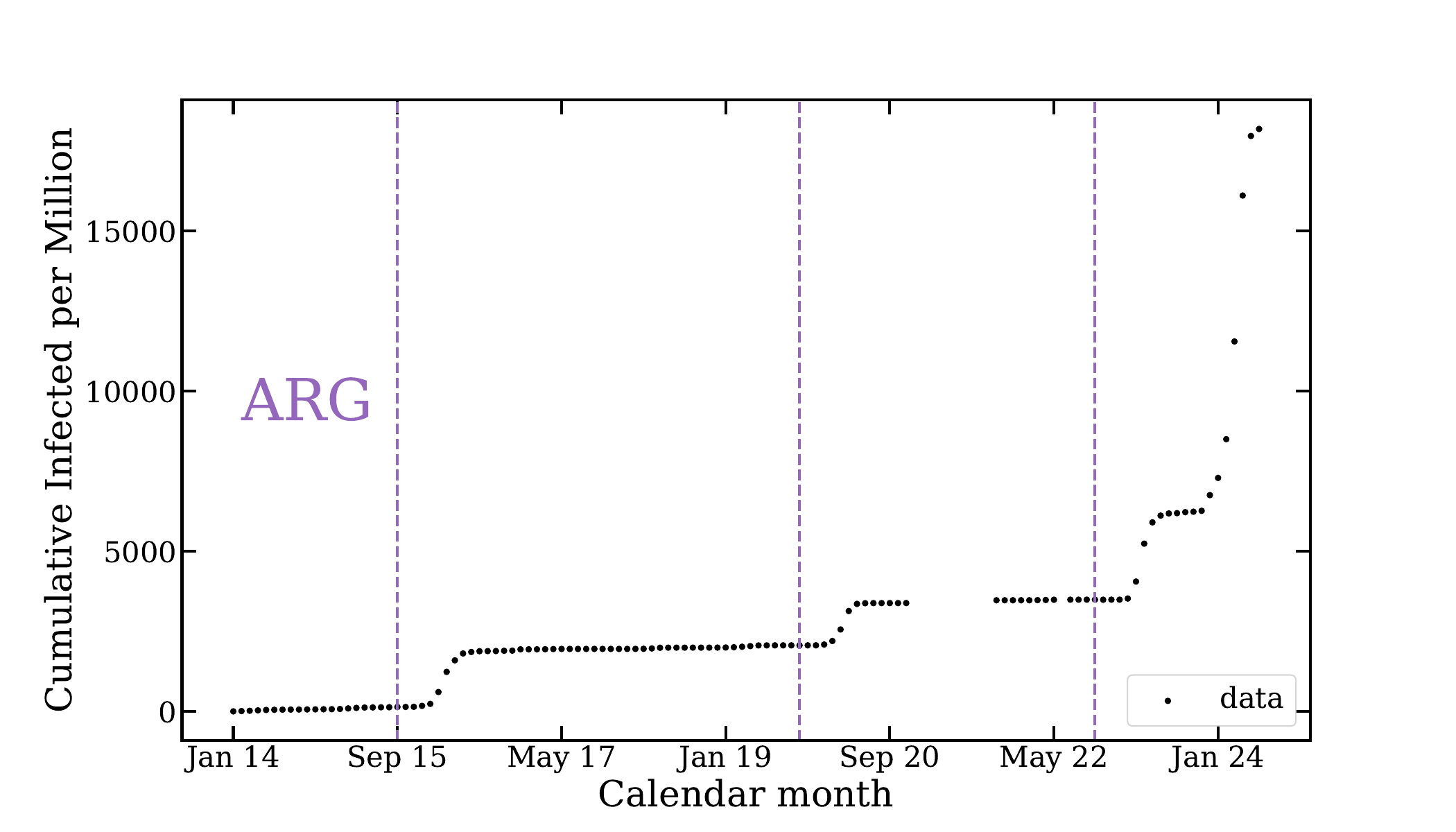}
		 \caption{ARG}
		 \label{fig:ARG}
	      \end{subfigure}
	     \begin{subfigure}{0.45\linewidth}
		 \includegraphics[width=\linewidth]{ 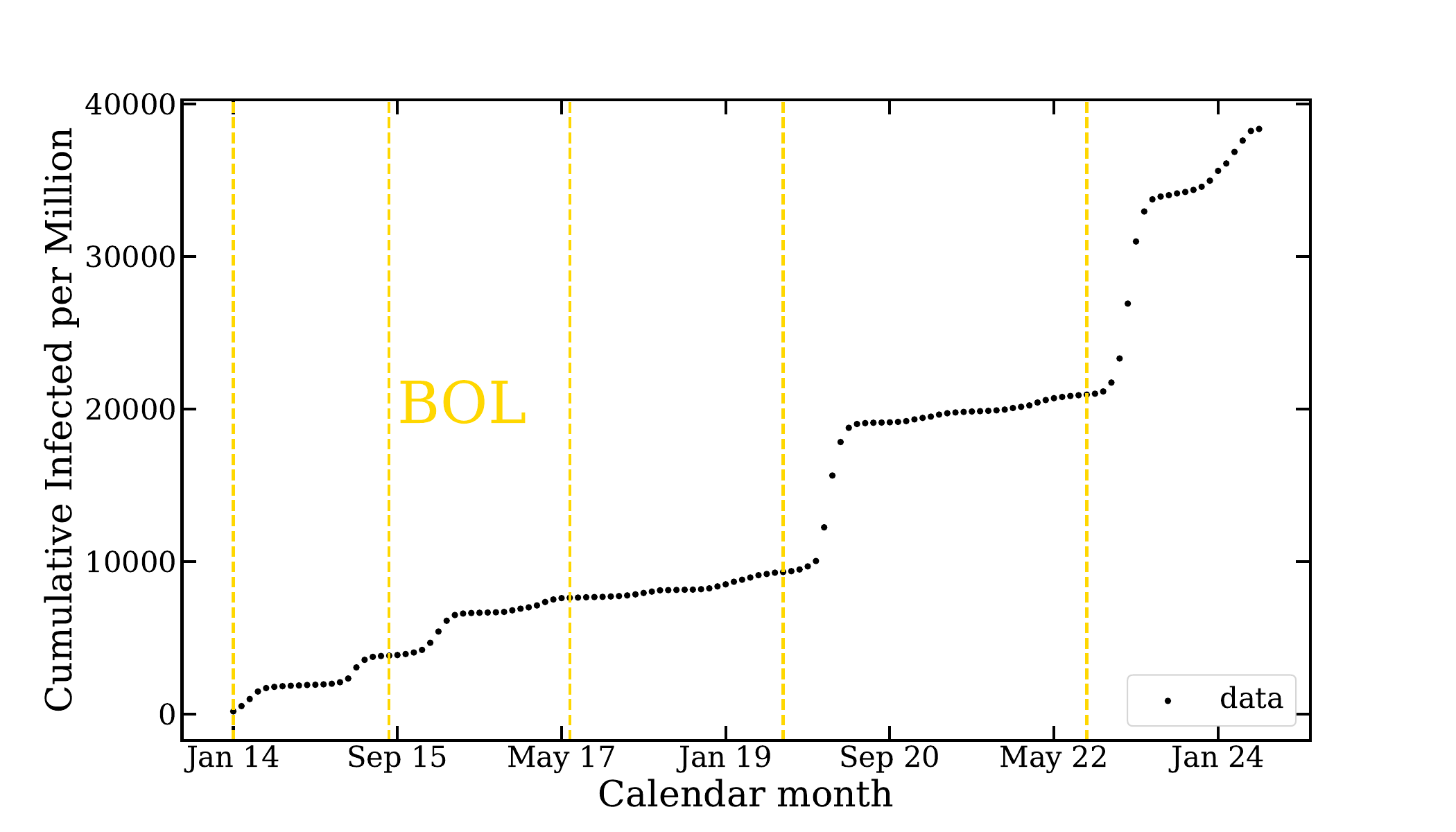}
		 \caption{BOL}
		 \label{fig:BOL}
	      \end{subfigure}
      % \label{fig:plotsPartVax}
               \vfill
        \medskip
            \begin{subfigure}{0.45\linewidth}
    		  \includegraphics[width=\linewidth]{ 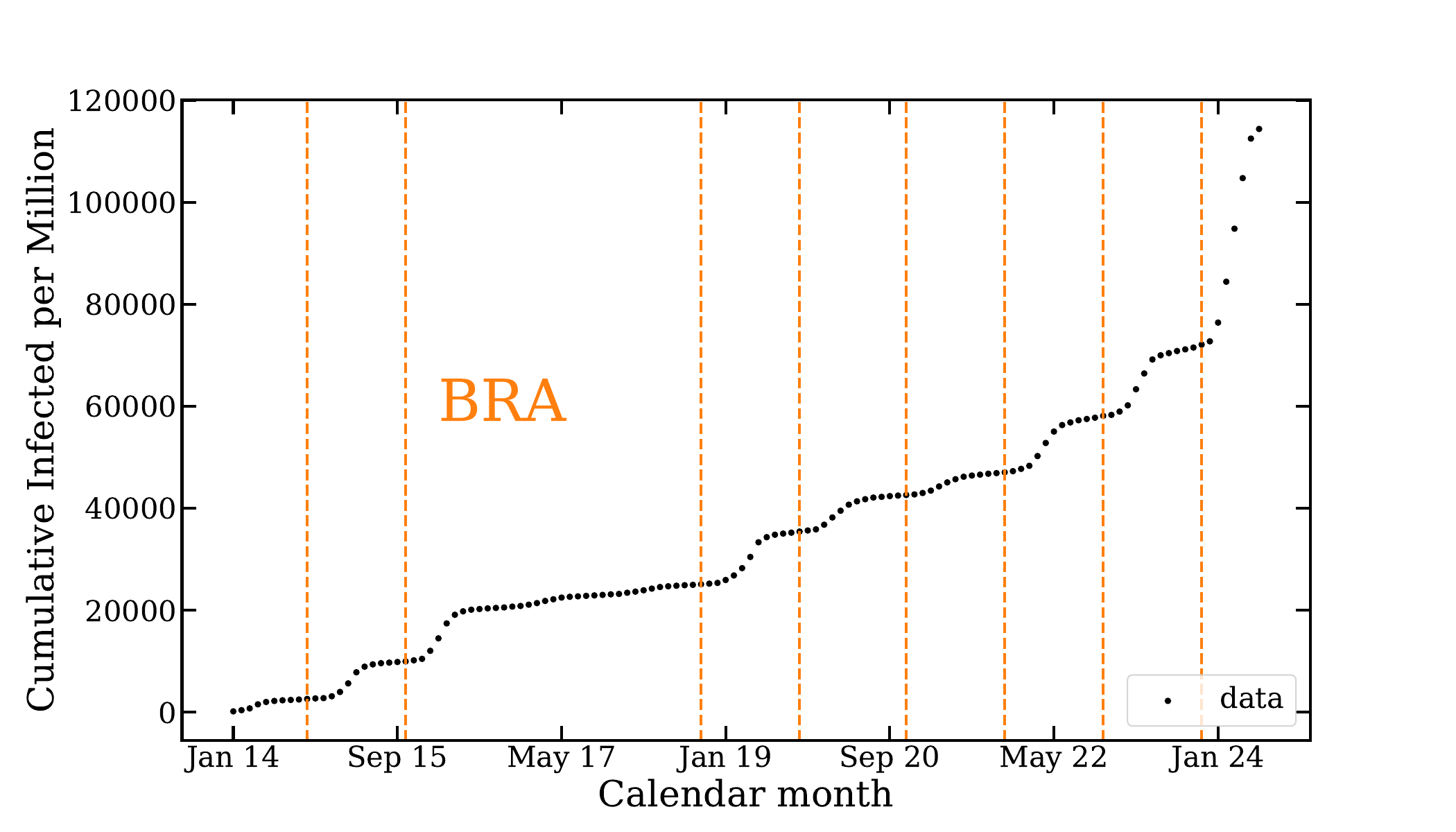}
    		  \caption{BRA}
    		  \label{fig:BRA}
    	       \end{subfigure}
    	     \begin{subfigure}{0.45\linewidth}
    		 \includegraphics[width=\linewidth]{ 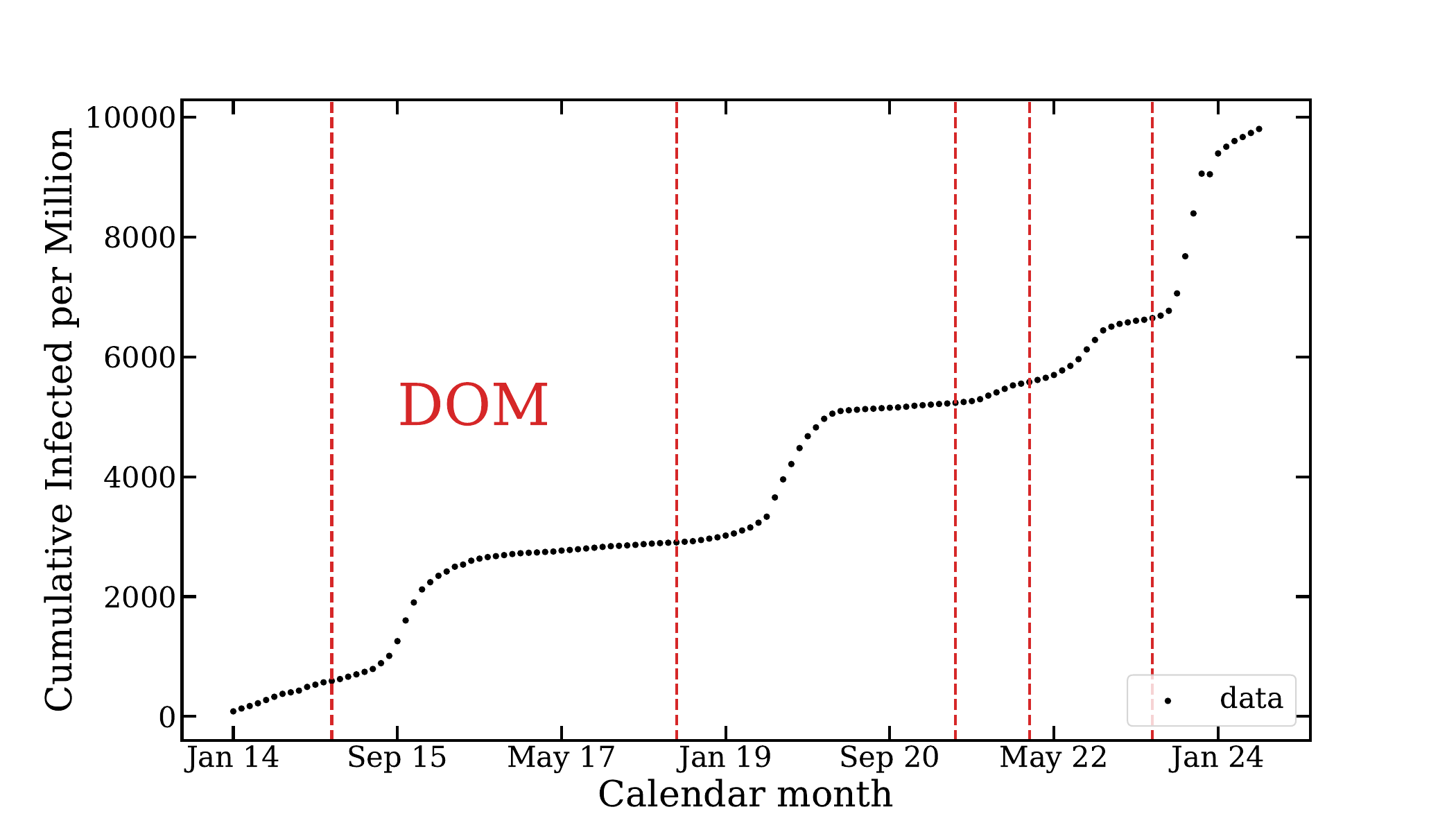}
    		 \caption{DOM}
    		 \label{fig:DOM}
    	      \end{subfigure}
    	  \vfill
        \medskip
            \begin{subfigure}{0.45\linewidth}
    		  \includegraphics[width=\linewidth]{ 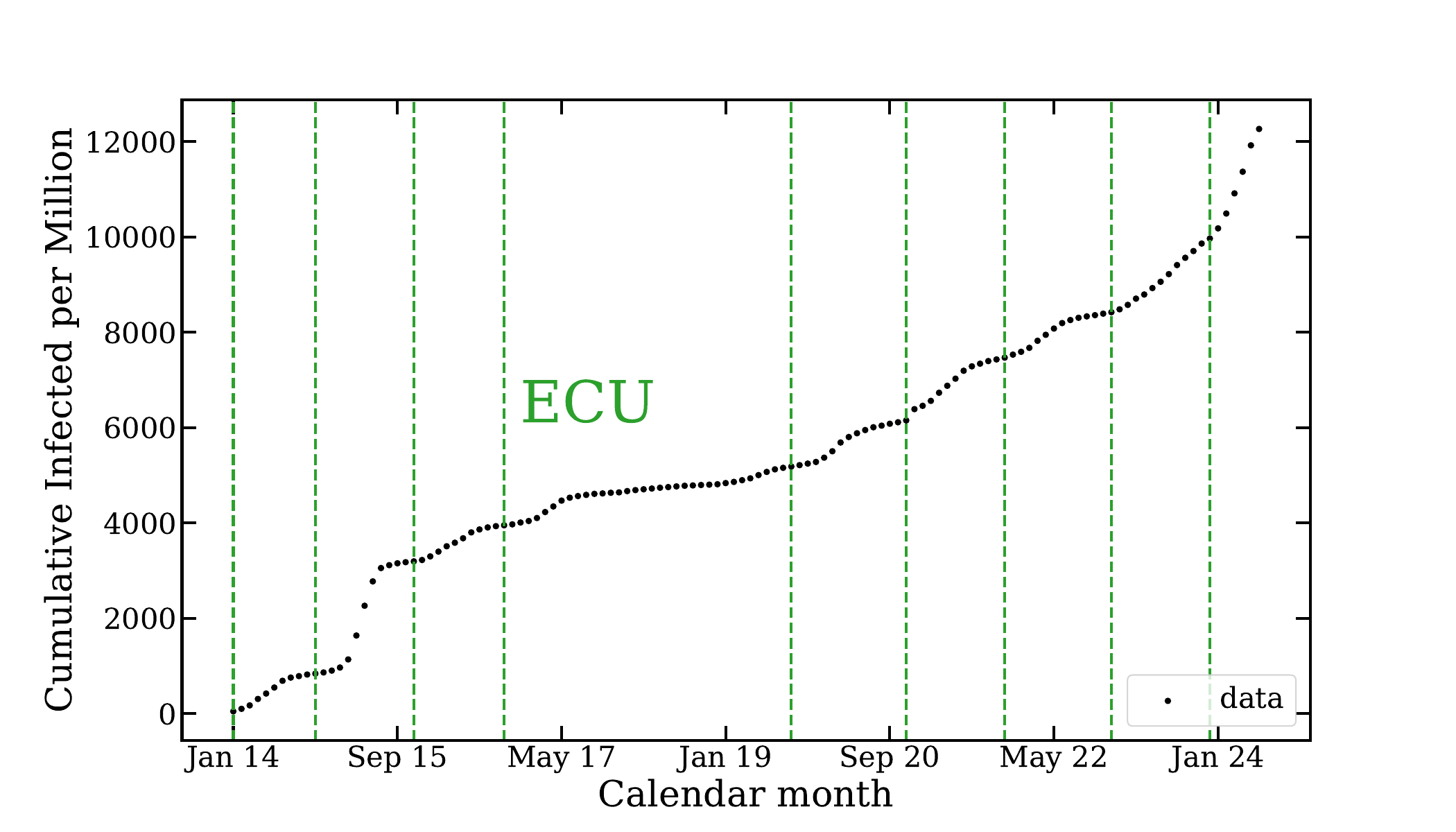}
    		  \caption{ECU}
    		  \label{fig:ECU}
    	       \end{subfigure}
    	     \begin{subfigure}{0.45\linewidth}
    		 \includegraphics[width=\linewidth]{ 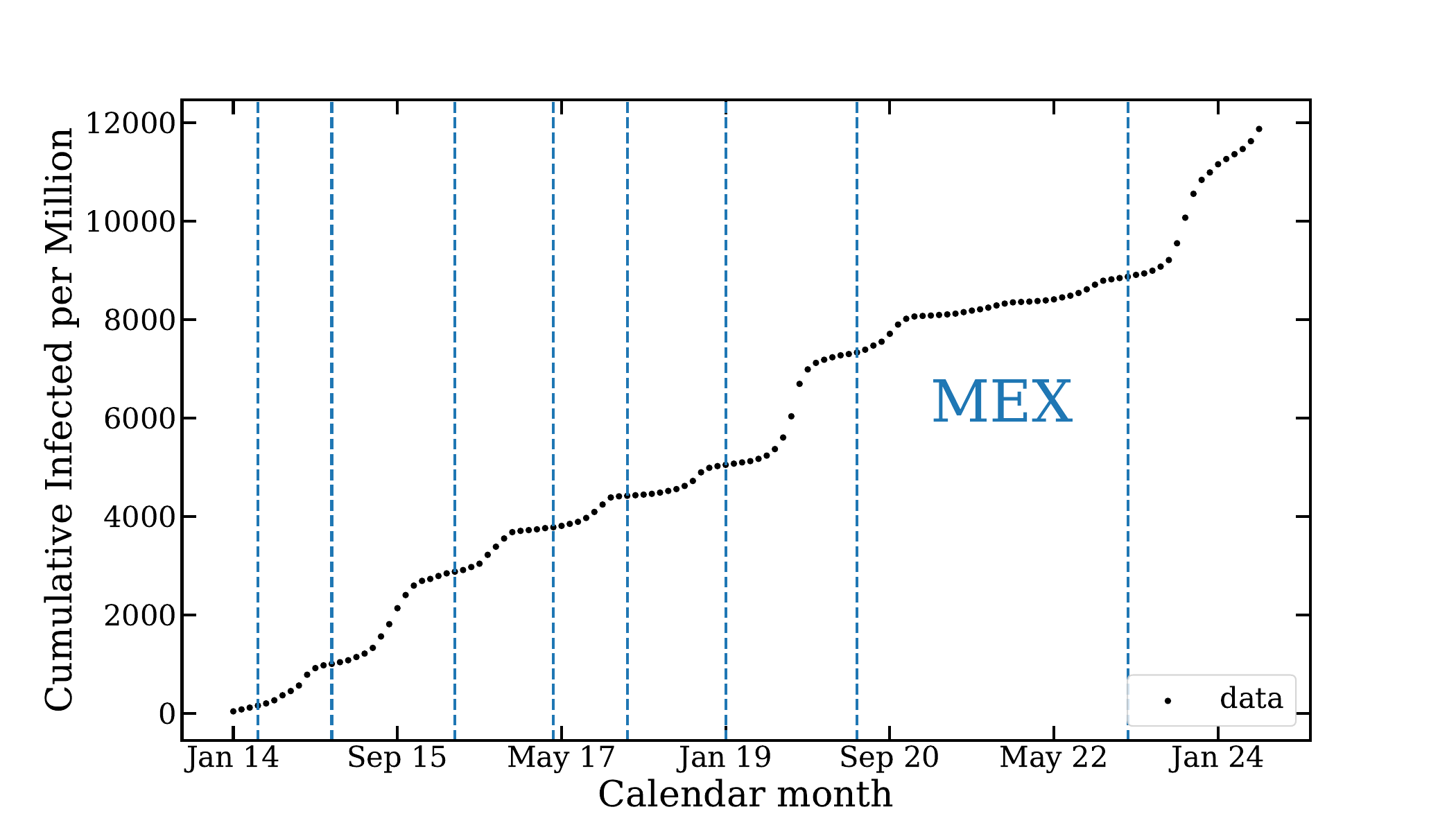}
    		 \caption{MEX}
    		 \label{fig:MEX}
    	      \end{subfigure}
    \caption{Plots of the official data of cumulative infected people per Million together with the eRG fit.}
    \label{fig:cumulative_ondate}
\end{figure}

We apply the eRG model to six Latin American countries: Argentina (ARG), Bolivia (BOL), Brazil (BRA), Ecuador (ECU), Dominican Republic (DOM) and Mexico (MEX). The latter are selected because public data on Dengue are available. Furthermore, they show a recognizable multiple wave pattern. To set the stage, we start by examining the official data obtained from the World Health Organization (WHO), accessible at \url{https://worldhealthorg.shinyapps.io/Dengue_global/}. Our initial analysis will provide a foundation for understanding the broader trends and patterns observed across countries. Specifically, in Figure \ref{fig:cumulative_ondate} we report the official data for cumulative infections, normalized by the population number in millions for each country. The vertical lines indicate the start of each pandemic wave.
%\FloatBarrier

\FloatBarrier

For each country and each wave, we applied the eRG model to describe the data. The solution of the model in Eq.\eqref{eq:beta0} suggests that the cumulative number of infected individuals is well represented by the following function 
\begin{equation}
    \mathcal{I}(t) =\exp{\frac{a e^{\gamma t}}{b+ e^{\gamma t}}}\ .
\end{equation}
The cumulative number of infected individuals grows rapidly at the beginning of the wave, then quickly approaches the plateau at later times. According to the set of official data at our disposal, we indicate with $t$ the time measured in months. The parameter $a$ determines the height of the plateau, while $\ln(b)$ dictates the offset in the time of the start of the spread of the disease. The infection rate parameter $\gamma$ controls the slope of the curve and is measured in inverse time units. The parameter $a$ inherently reflects the number of infectious transmissions from the next neighbor, and is influenced by containment measures and the size of the population.

\begin{figure}[h!]
      \centering
	     \begin{subfigure}{0.45\linewidth}
		 \includegraphics[width=\linewidth]{ 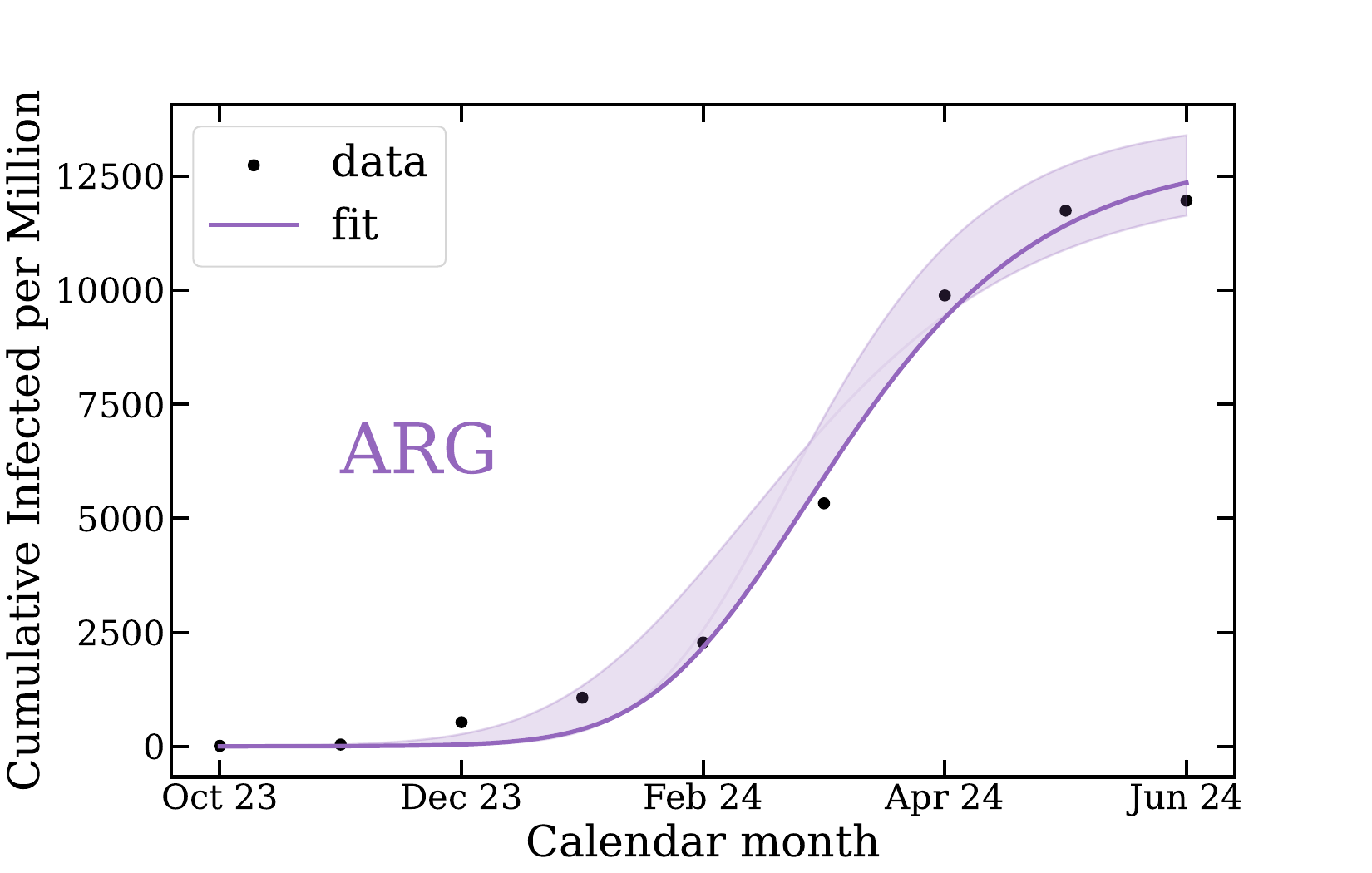}
		 \caption{ARG}
		 \label{fig:ARGFIT}
	      \end{subfigure}
	     \begin{subfigure}{0.45\linewidth}
		 \includegraphics[width=\linewidth]{ 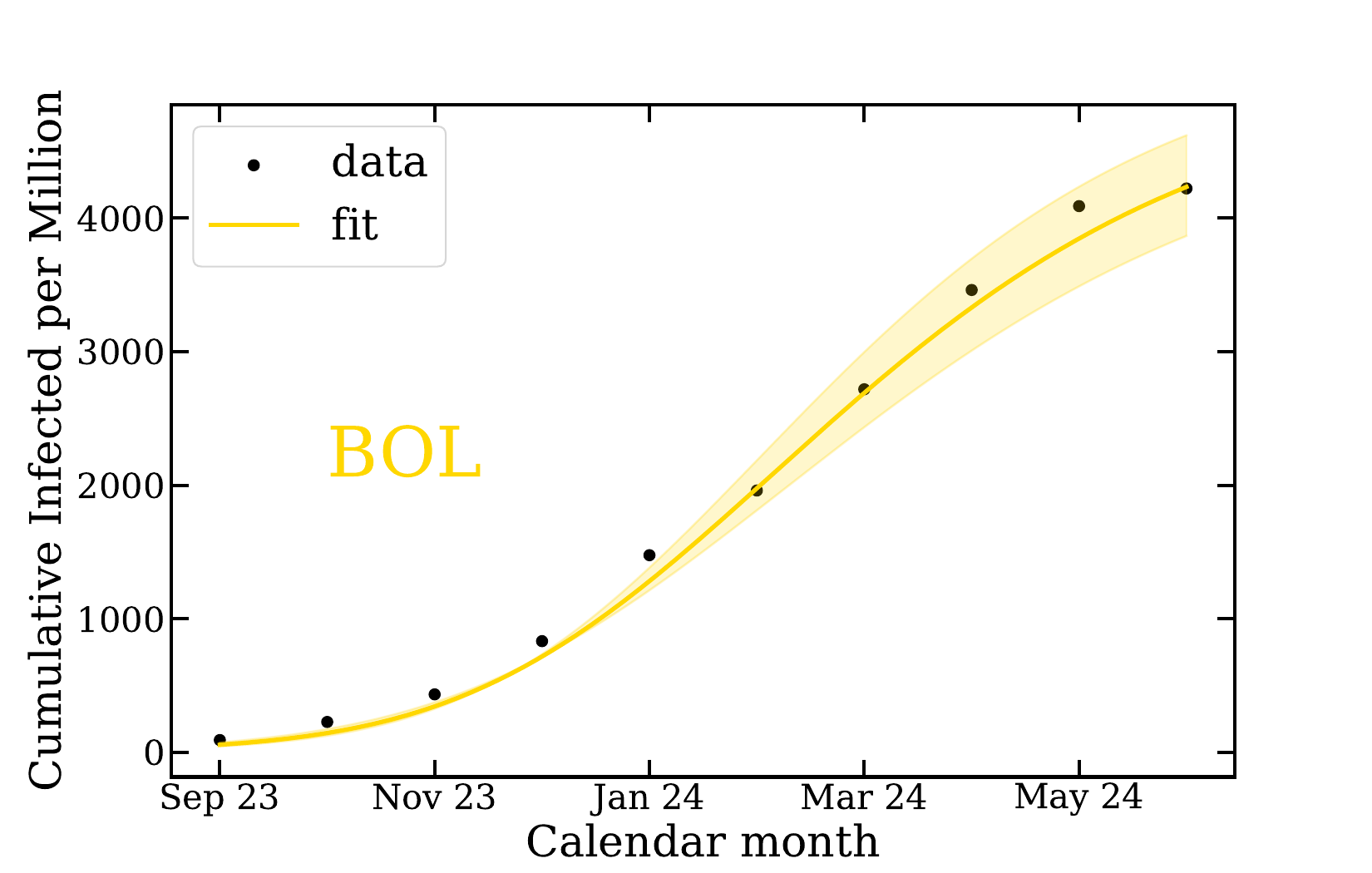}
		 \caption{BOL}
		 \label{fig:BOLFIT}
	      \end{subfigure}
      % \label{fig:plotsPartVax}
        \begin{subfigure}{0.45\linewidth}
    		  \includegraphics[width=\linewidth]{ 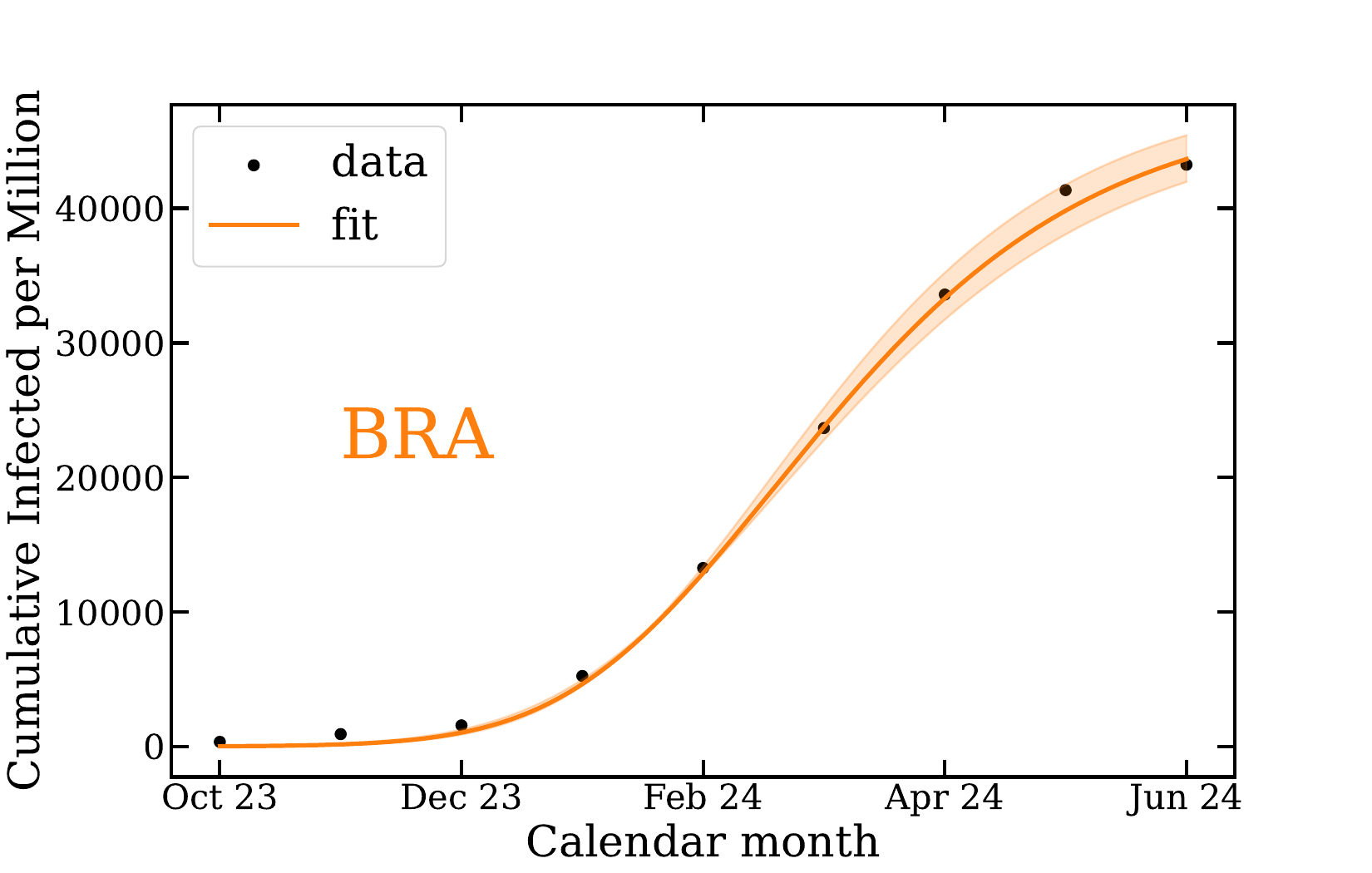}
    		  \caption{BRA}
    		  \label{fig:BRAFIT}
    	       \end{subfigure}
            \begin{subfigure}{0.45\linewidth}
    		  \includegraphics[width=\linewidth]{ 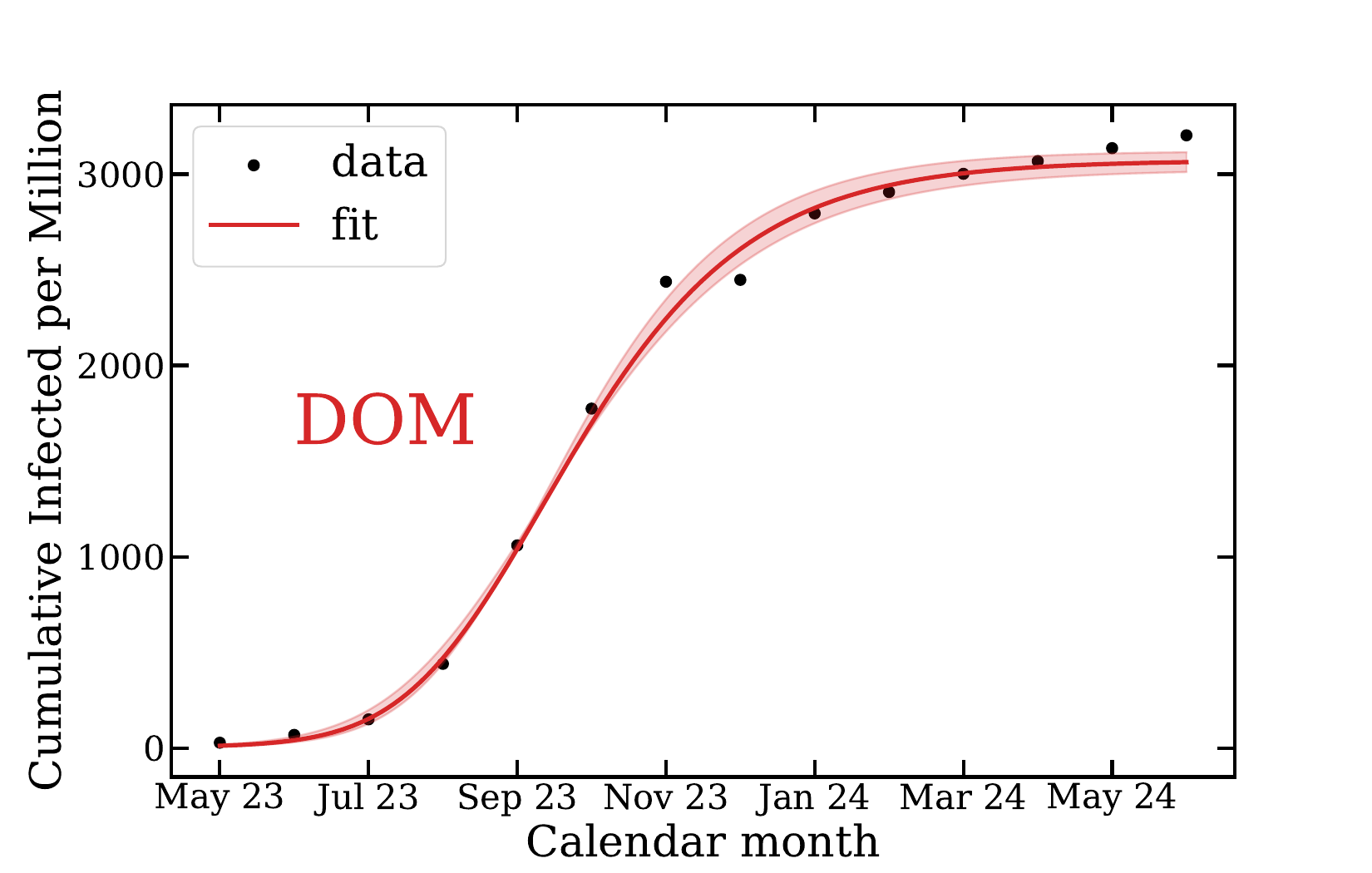}
    		  \caption{DOM}
    		  \label{fig:DOMFIT}
    	       \end{subfigure}
             \vfill
        \medskip
    	     \begin{subfigure}{0.45\linewidth}
    		 \includegraphics[width=\linewidth]{ 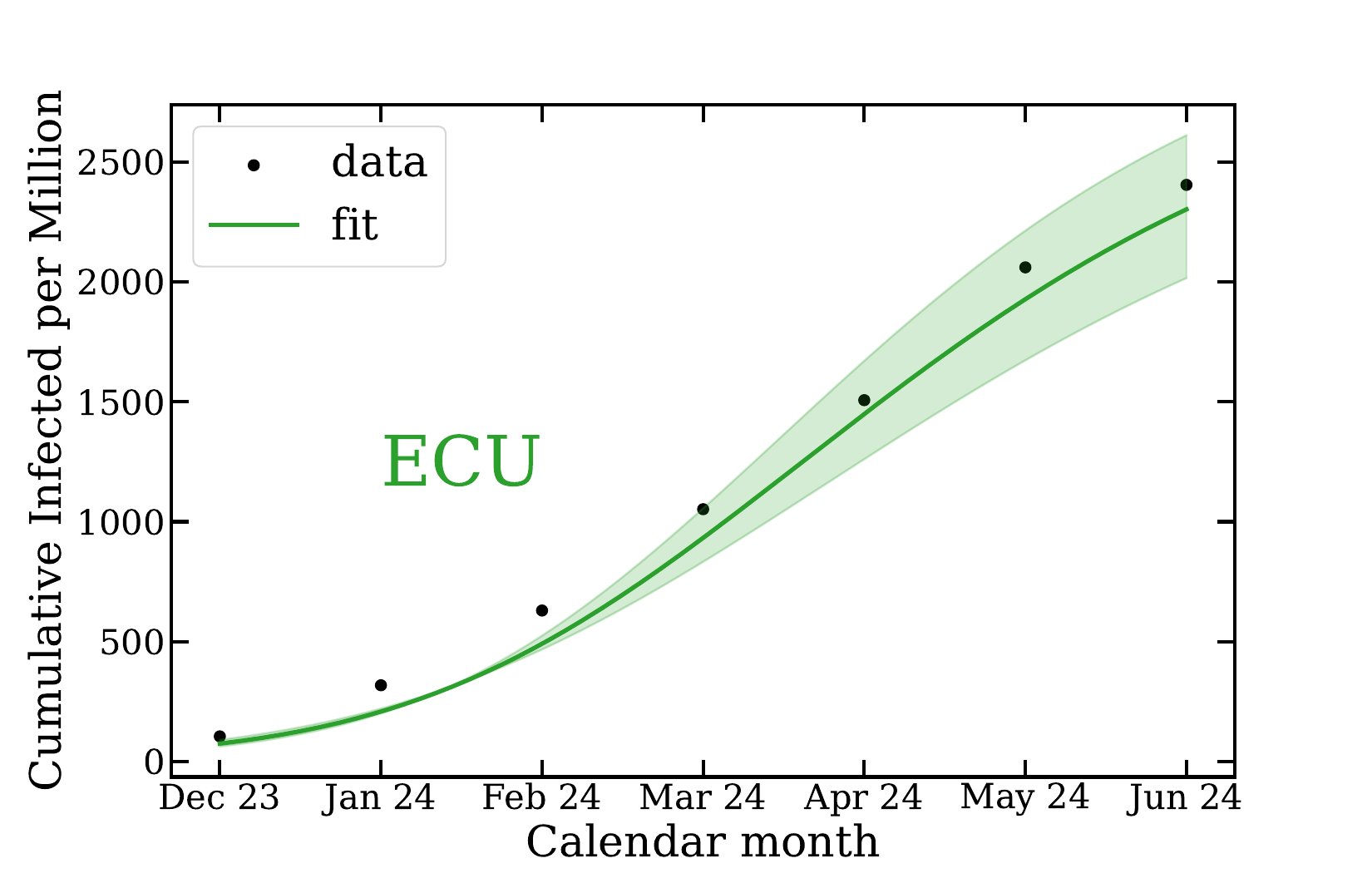}
    		 \caption{ECU}
    		 \label{fig:ECUFIT}
    	      \end{subfigure}
            \begin{subfigure}{0.45\linewidth}
    		  \includegraphics[width=\linewidth]{ 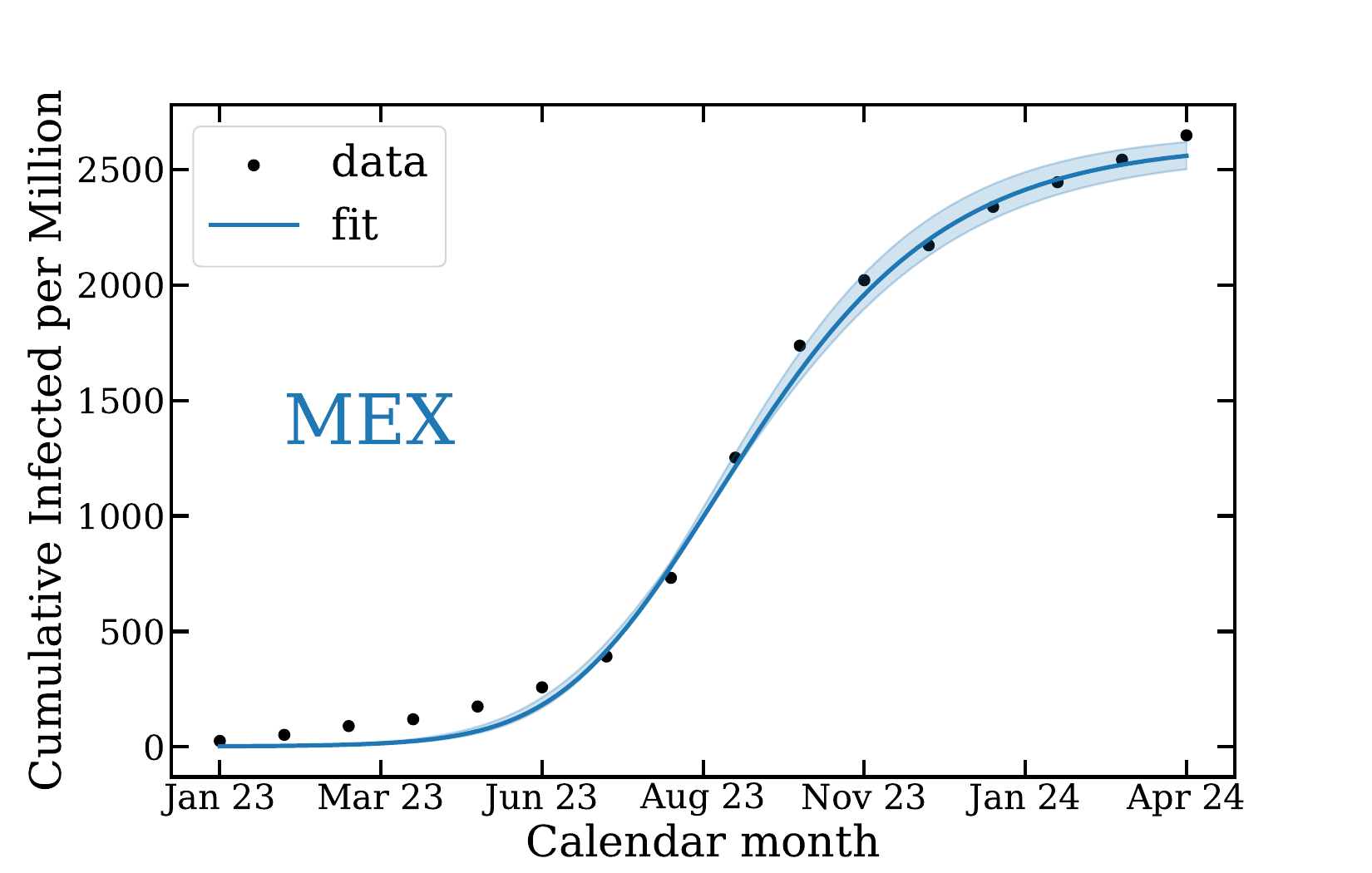}
    		  \caption{MEX}
    		  \label{fig:MEXFIT}
    	       \end{subfigure}
    \caption{Plots of the official data of cumulative infected people per Million together with the eRG fit for the last wave.}
    \label{fig:cumulative_last_wave}
\end{figure}

To illustrate the power of the approach, we present in Figure~\ref{fig:cumulative_last_wave} the model results describing the evolution of the cumulative number of infected individuals limited to the last pandemic wave. The straight lines represent the best-fit eRG curve while the bands correspond to one-sigma confidence interval. In Appendix \ref{tables} we report in Tables \ref{tab:parameters_ARG}--\ref{tab:parameters_NPL} the best-fit results for the eRG parameters corresponding to all the pandemic waves for each country.

\FloatBarrier

\FloatBarrier
As it is clear from the analysis the eRG framework provides a satisfactory modeling of the pandemic waves already at one sigma level.

\FloatBarrier

 We now turn to a more detailed presentation of the results, examining each country individually. This country-by-country analysis allows us to achieve deeper understandings of the trends, challenges, and outcomes observed across different regions. The Dengue virus represents a social alarm in the Latin America with the actual risks and danger varying across the various countries due to a number of socio-economic factors. \\
\paragraph{Argentina}
By closely examining Figure \ref{fig:cumulative_ondate} we observe that the peak of each wave consistently occurs around April and May. Consequently, to analyze the most recent wave of the disease, we selected October 2023 as the starting point for this new wave. The results, shown in Figure \ref{fig:cumulative_last_wave}, illustrate the number of infected cases over time, along with the best-fit curve, which yields 
\begin{equation}
  a =  9.47 \pm 0.06,\qquad b =9.92 \pm 6.60 \ ,  \qquad \gamma = 0.94 \pm 0.15\ . 
\end{equation}
It is reasonable to expect that a new wave will start soon, during the Fall 2024 season.
\paragraph{Bolivia}
In Bolivia, the last Dengue wave begins in September 2023. The outcome, displayed in Figure \ref{fig:cumulative_last_wave}, provides a timeline of infected cases, along with a best-fit model that accurately represents the observed pattern with the following values
\begin{equation}
  a =  8.53 \pm 0.05,\qquad b =  1.10 \pm 0.15 \ , \qquad \gamma = 0.44 \pm 0.04\ . 
\end{equation}

\paragraph{Brazil}
As shown in Figure \ref{fig:cumulative_ondate}, the peaks of the previous waves are generally reached around February and March. To capture the onset of the latest wave, we commenced our study from October 2023. The outcome, displayed in Figure \ref{fig:cumulative_last_wave}, provides a timeline of infected cases, along with a best-fit model that accurately represents the observed pattern with the following values
\begin{equation}
  a =  10.78 \pm 0.03,\qquad b =  2.18 \pm 0.33 \ , \qquad \gamma = 0.69 \pm 0.04\ . 
\end{equation}

\paragraph{Dominican Republic}
For this country, to ensure a thorough analysis of the most recent wave, we start from May 2023. The results, as shown in Figure \ref{fig:cumulative_last_wave}, detail the number of infected cases over time, paired with a best-fit curve that very well captures the trend. We obtain the following set of fitted parameters
\begin{equation}
  a =  8.03 \pm 0.02,\qquad b =2.29\pm 0.50 \ ,  \qquad \gamma = 0.67 \pm 0.05\ . 
\end{equation}

\paragraph{Ecuador}
In Ecuador, the most recent Dengue wave begins in December 2023. Figure \ref{fig:cumulative_ondate} illustrates the progression of infection cases over time, along with a best-fit curve that aligns closely with the data. The fitted parameters obtained from this analysis are
\begin{equation}
  a =  8.01 \pm 0.08,\qquad b = 0.85 \pm 0.09 \ , \qquad \gamma = 0.54 \pm 0.05\ . 
\end{equation}

\paragraph{Mexico}
 The fitted parameters obtained from the analysis of the most recent Dengue wave for Mexico, that we establish to start in January 2023, are
\begin{equation}
  a =  7.87 \pm 0.02,\qquad b = 6.86 \pm 1.53 \ , \qquad \gamma = 0.52 \pm 0.03\ , 
\end{equation}
that lead to the best-fit blue curve shown in Figure \ref{fig:cumulative_last_wave}.

\begin{figure}[h!]
      \centering
	     \begin{subfigure}{0.45\linewidth}
		 \includegraphics[width=\linewidth]{ 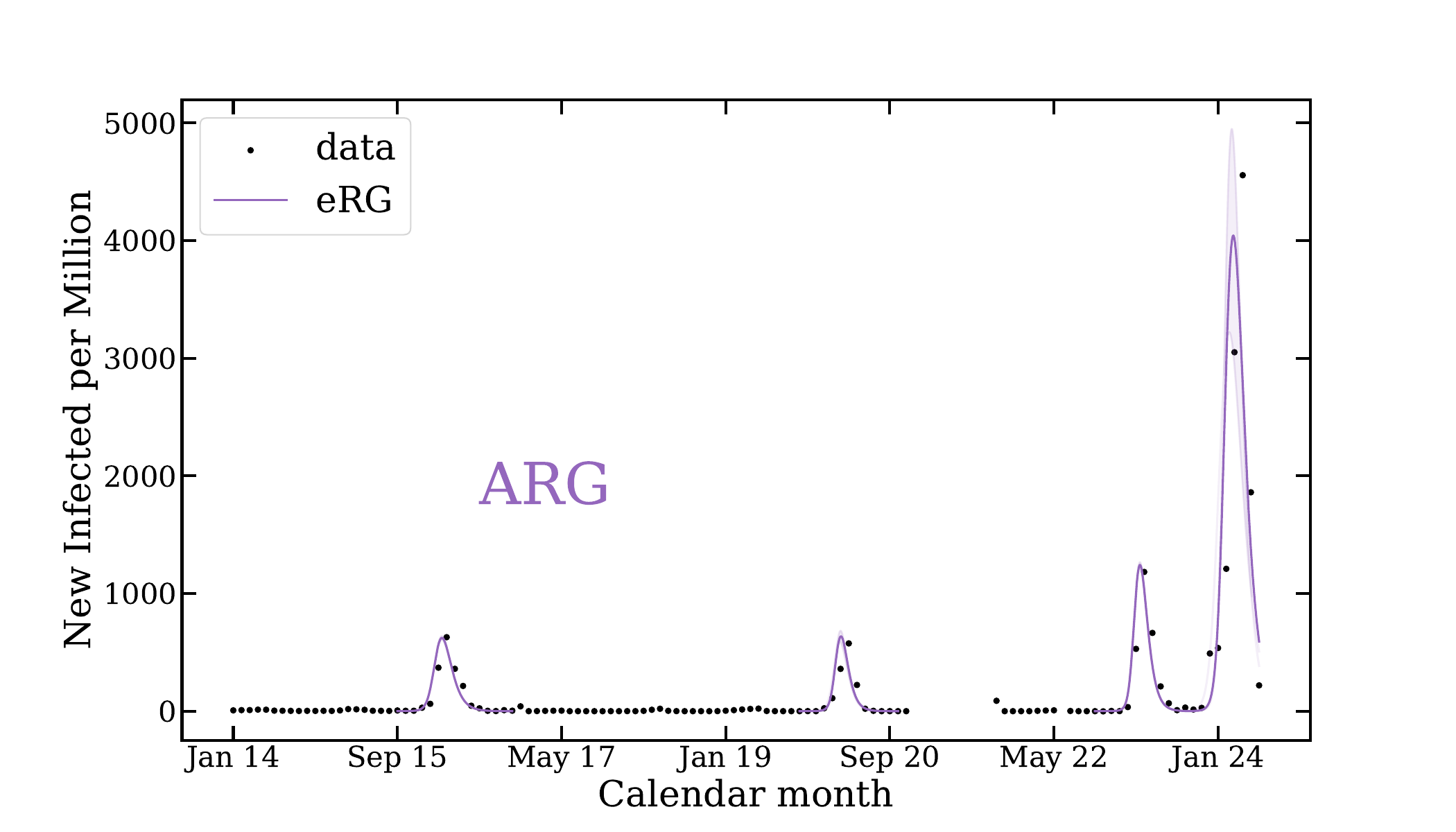}
		 \caption{ARG}
		 \label{fig:ARGni}
	      \end{subfigure}
	     \begin{subfigure}{0.45\linewidth}
		 \includegraphics[width=\linewidth]{ 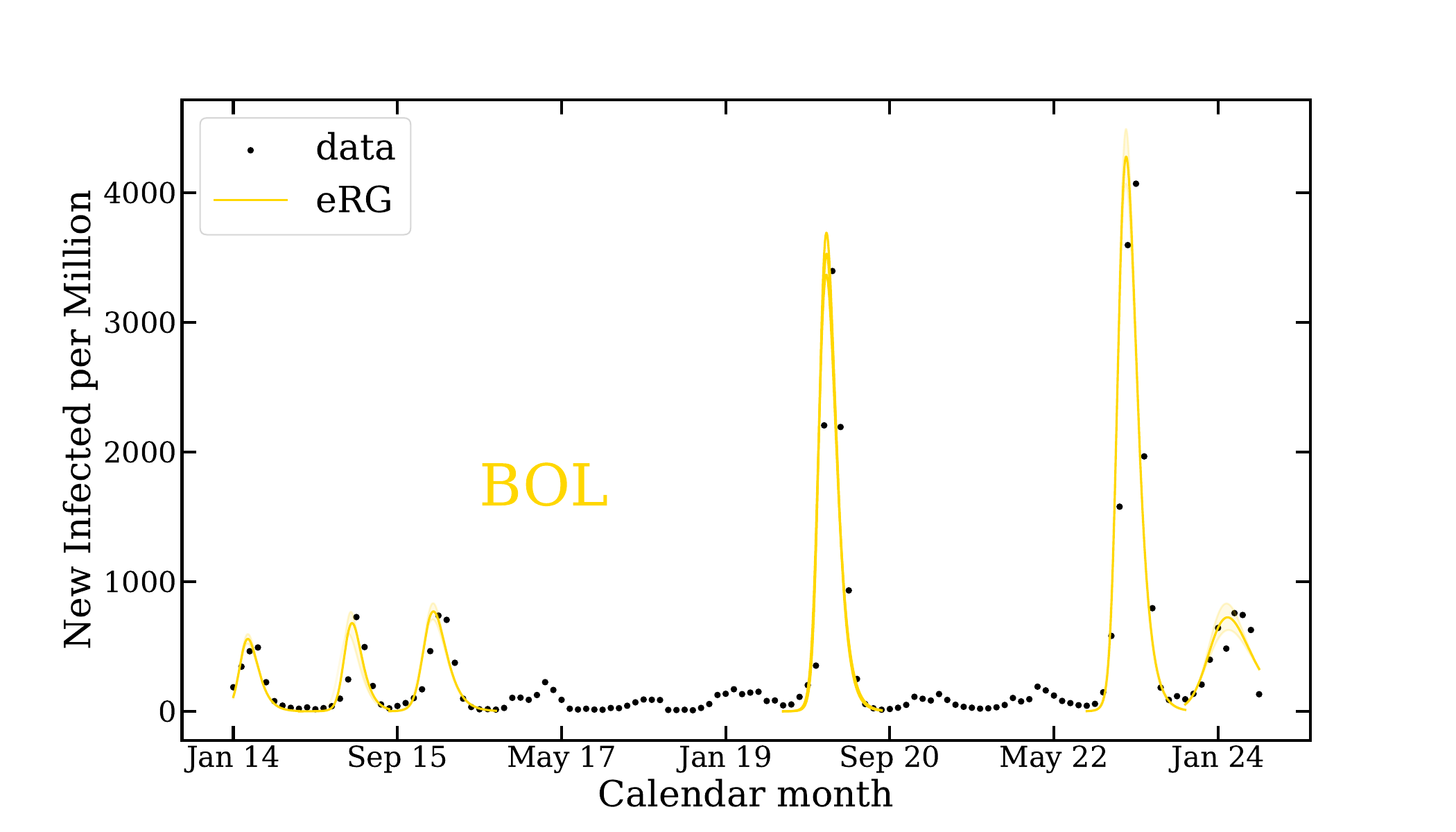}
		 \caption{BOL}
		 \label{fig:BOLni}
	      \end{subfigure}
      % \label{fig:plotsPartVax}
               \vfill
        \medskip
         \begin{subfigure}{0.45\linewidth}
    		  \includegraphics[width=\linewidth]{ 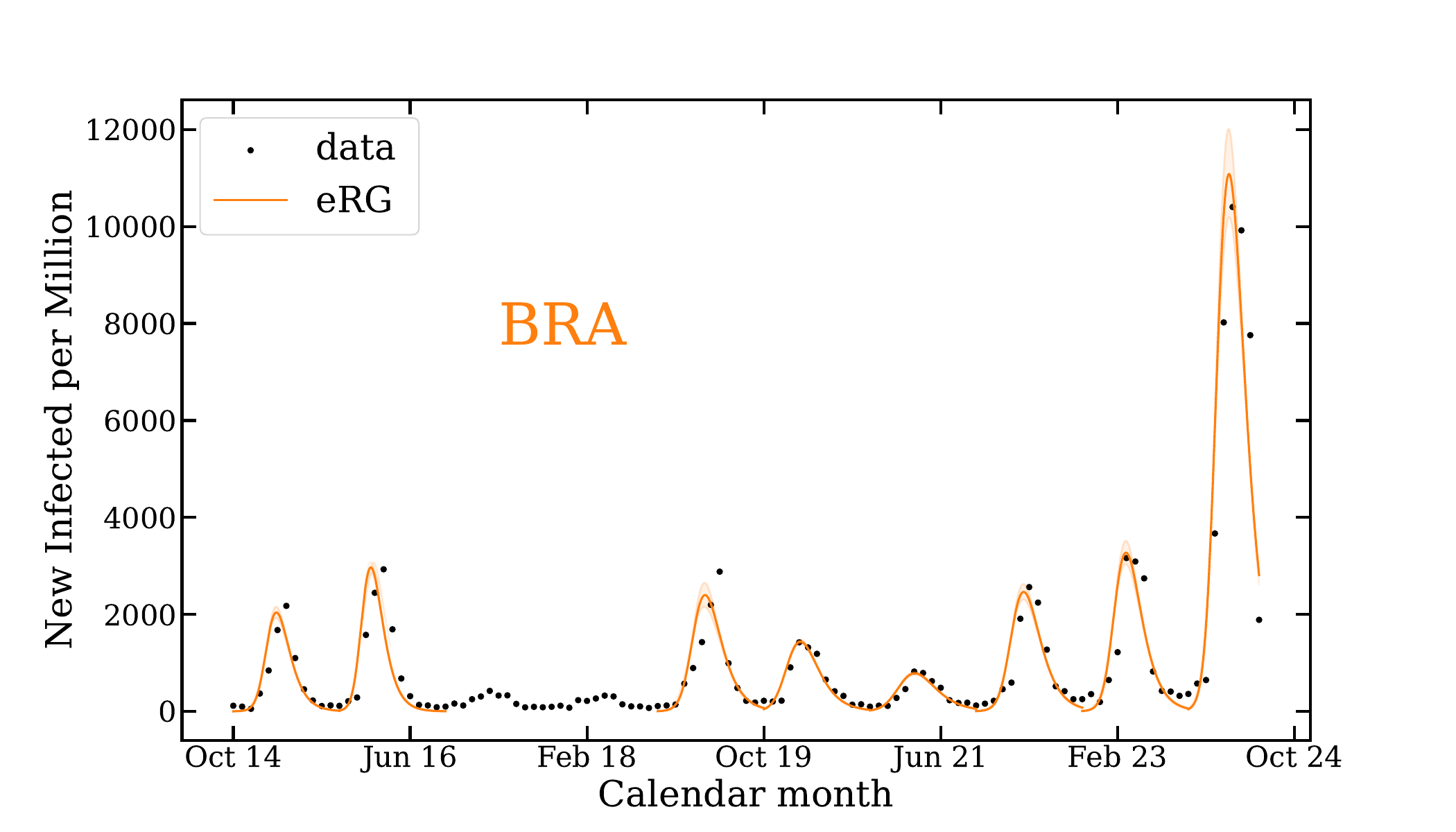}
    		  \caption{BRA}
    		  \label{fig:BRAni}
    	       \end{subfigure}
            \begin{subfigure}{0.45\linewidth}
    		  \includegraphics[width=\linewidth]{ 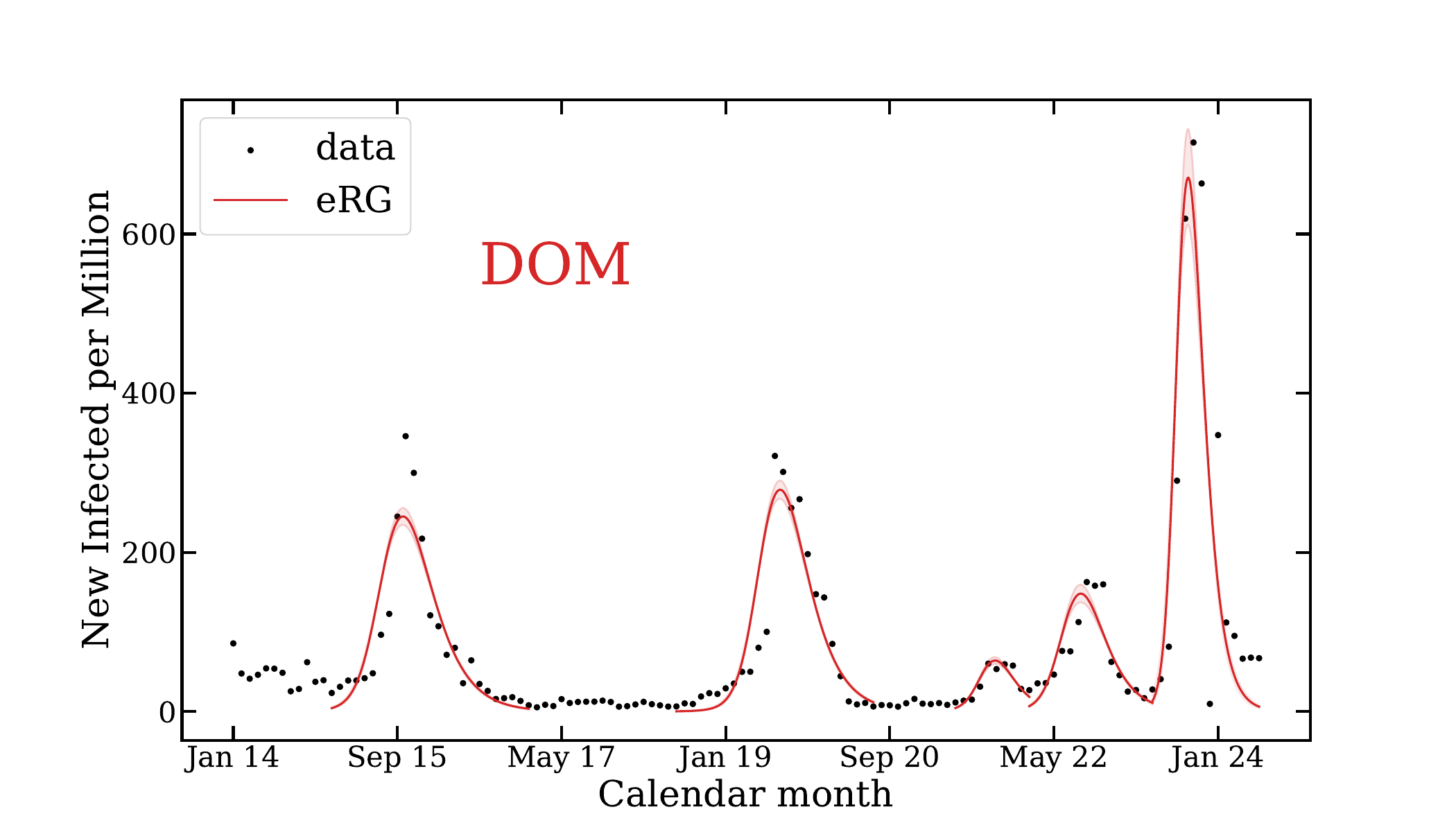}
    		  \caption{DOM}
    		  \label{fig:DOMni}
    	       \end{subfigure}
            \vfill
        \medskip
    	      \begin{subfigure}{0.45\linewidth}
    		  \includegraphics[width=\linewidth]{ 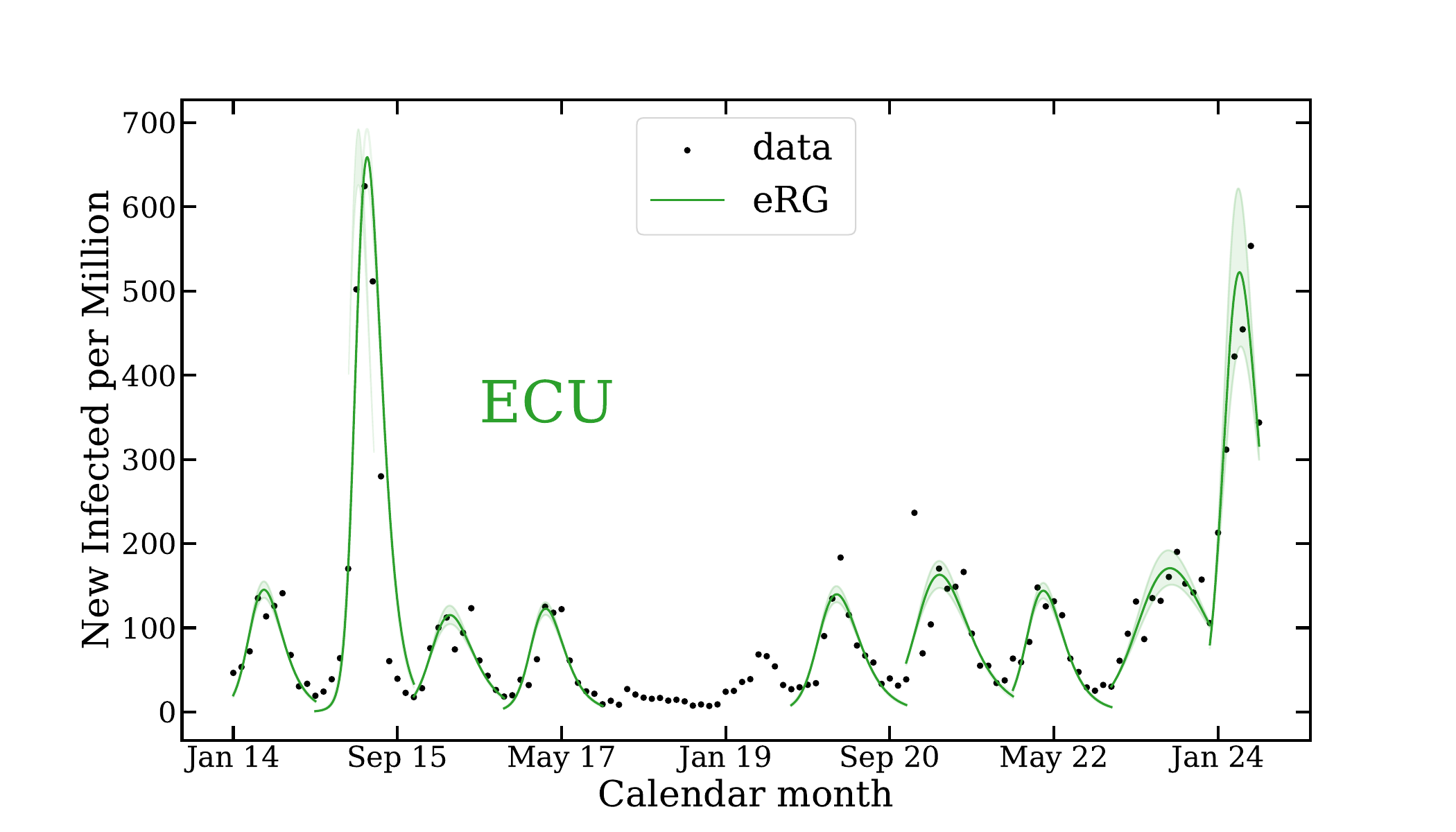}
    		  \caption{ECU}
    		  \label{fig:ECUni}
    	       \end{subfigure}
            \begin{subfigure}{0.45\linewidth}
    		  \includegraphics[width=\linewidth]{ 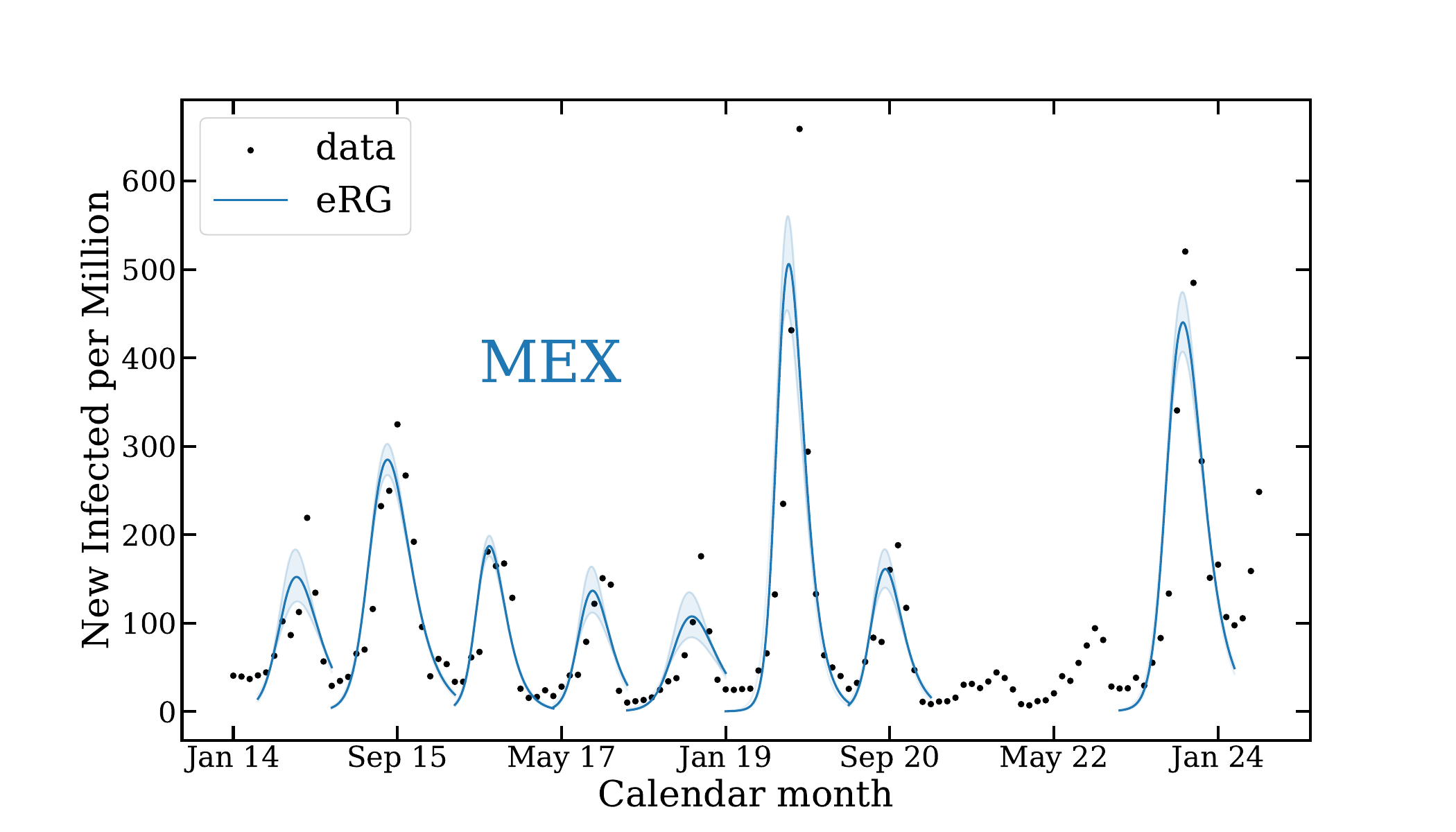}
    		  \caption{MEX}
    		  \label{fig:MEXni}
    	       \end{subfigure}
    \caption{Plots of the official data of new infected people per Million together with the eRG fit.}
    \label{fig:new_ondate}
\end{figure}

\vskip 2em
The result of a similar analysis is shown in Figure~ \ref{fig:new_ondate}, which illustrates the fit of the eRG model to new infections for all the pandemic waves for each country. For most countries, data on new infections show a significant decrease during 2020 and 2021, which we attribute to the impact of the COVID-19 pandemic. This decline likely reflects both reduced exposure to the virus and decreased opportunities for clinical testing of infected individuals. Consequently, the overall result indicates a gap in the dataset recorded during 2020, highlighting the challenges in accurately capturing infection rates during this period.
\FloatBarrier

\FloatBarrier

It is clear from the analysis of the Latin American countries that the eRG offers an excellent way to model the data globally. Encouraged by the results we now move to investigate the Dengue pandemic in Asia.

\subsection{Asia}
\begin{figure}[h!]
      \centering
       \begin{subfigure}{0.45\linewidth}
		 \includegraphics[width=\linewidth]{ 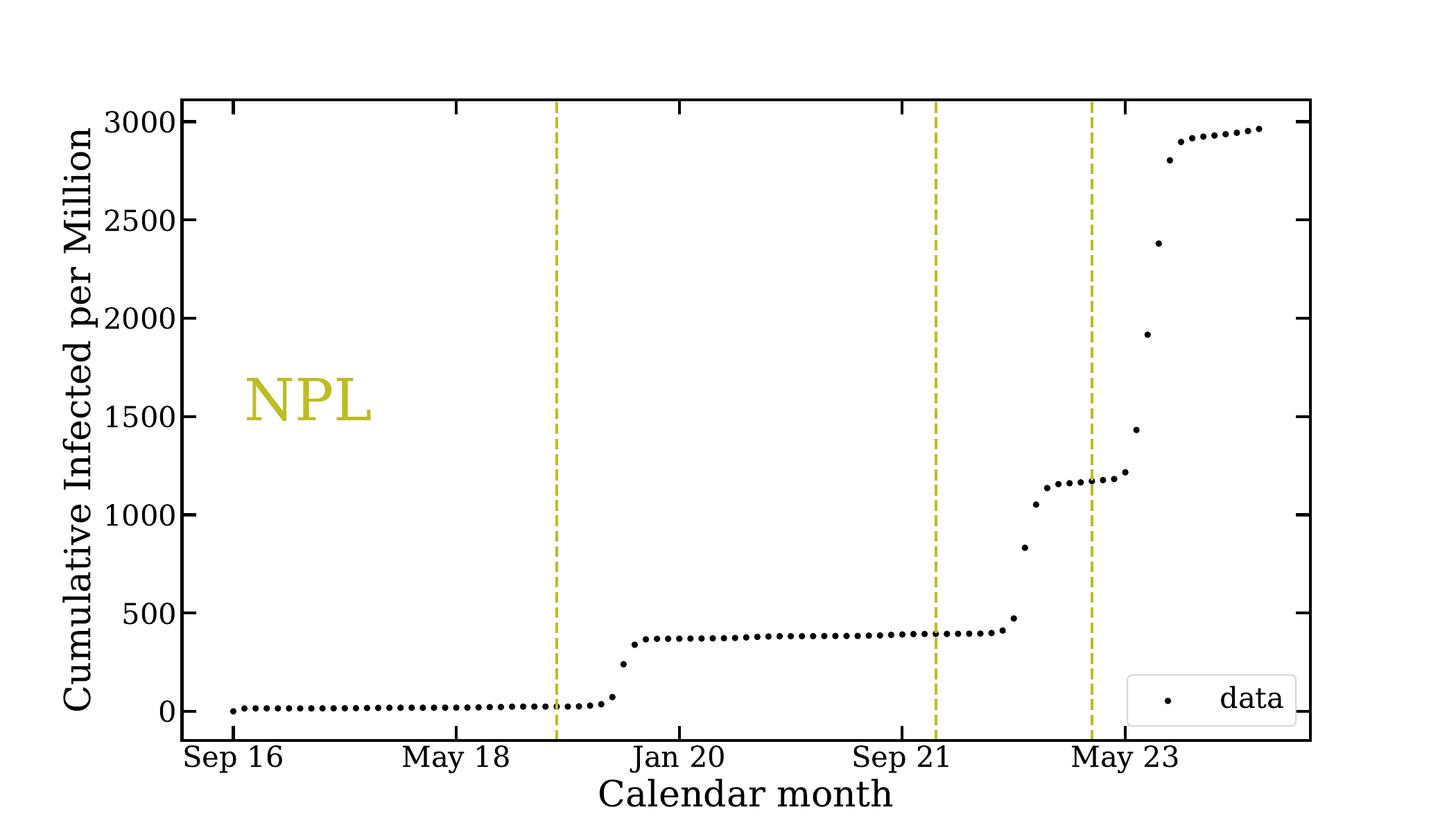}
		 \caption{NPL}
		 \label{fig:NPL}
	      \end{subfigure}
	\begin{subfigure}{0.52\linewidth}
		 \includegraphics[width=\linewidth]{ 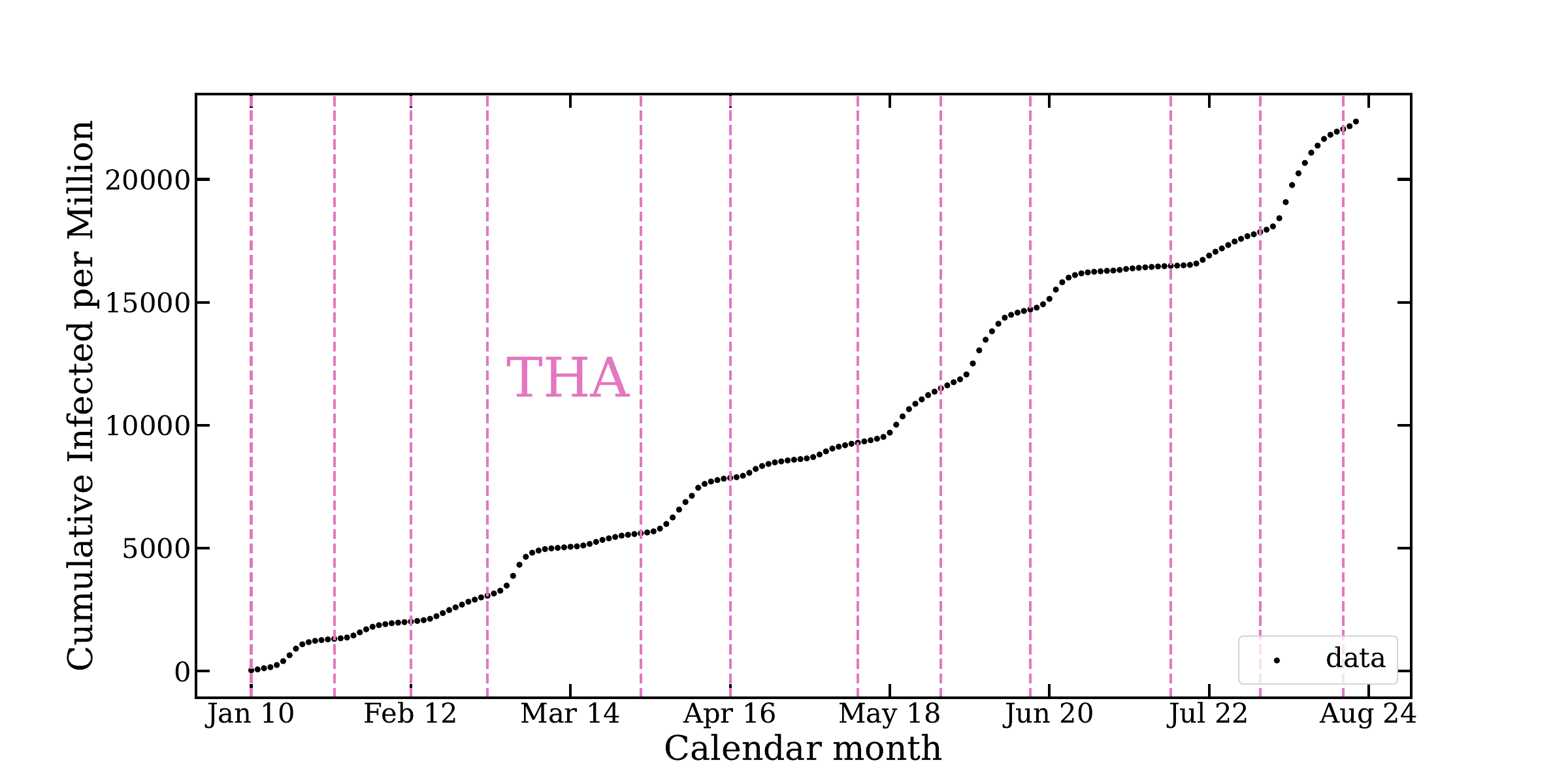}
		 \caption{THA}
		 \label{fig:THA}
	      \end{subfigure}
    \caption{Plots of the official data of cumulative infected people per Million together with the eRG fit.}
    \label{fig:cumulative_asia}
\end{figure}

Following the steps above, we investigate Dengue in two Asian countries with available data: Nepal and Thailand. Again, we examine the official cumulative number of infected individuals since the beginning of the corresponding epidemic outbreak, as shown in Figure \ref{fig:cumulative_asia}. 
\FloatBarrier

\begin{figure}[h!]
      \centering
       \begin{subfigure}{0.45\linewidth}
		 \includegraphics[width=\linewidth]{ 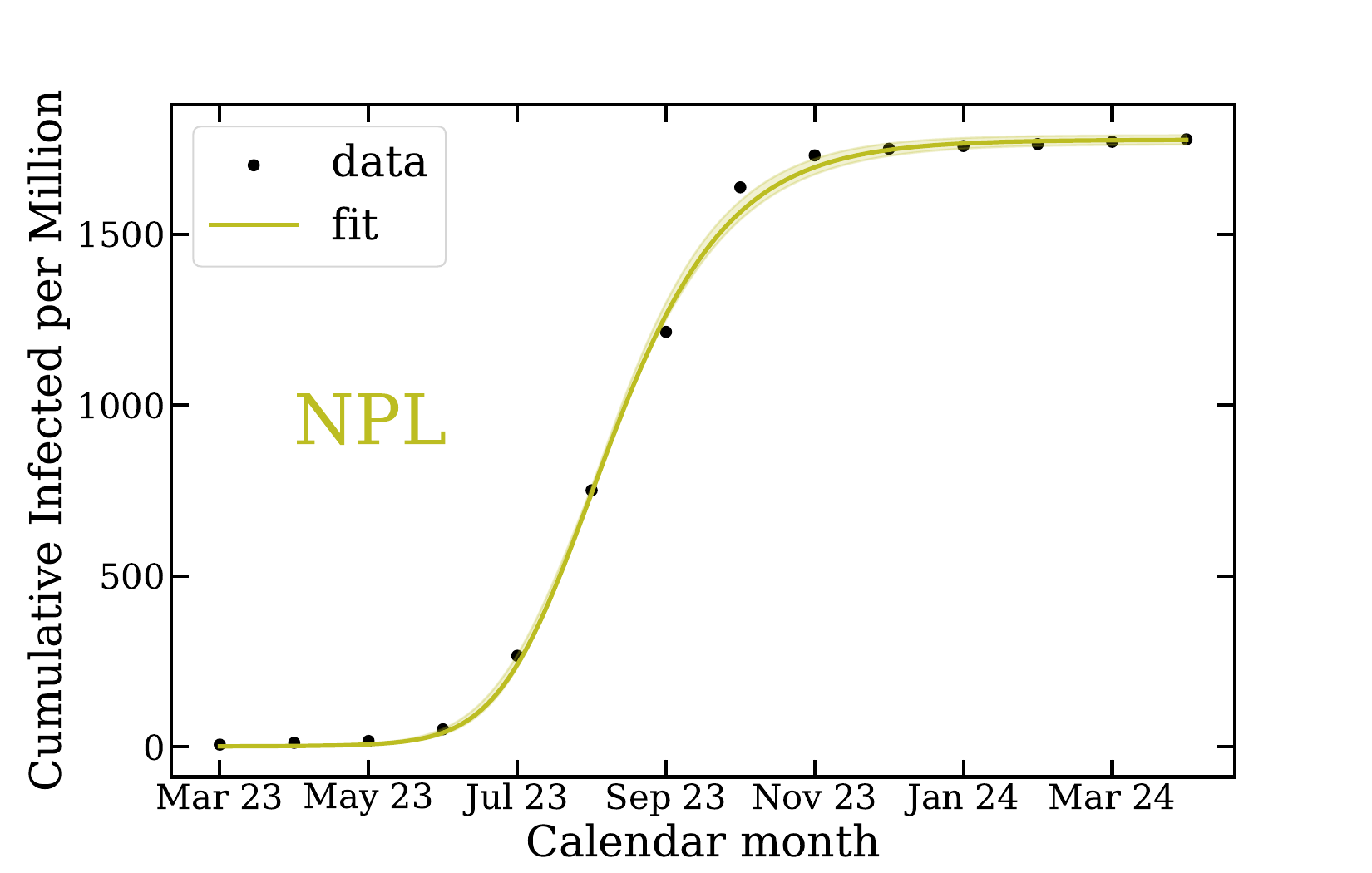}
		 \caption{NPL}
		 \label{fig:NPLFIT}
	      \end{subfigure}
	 \begin{subfigure}{0.45\linewidth}
		 \includegraphics[width=\linewidth]{ 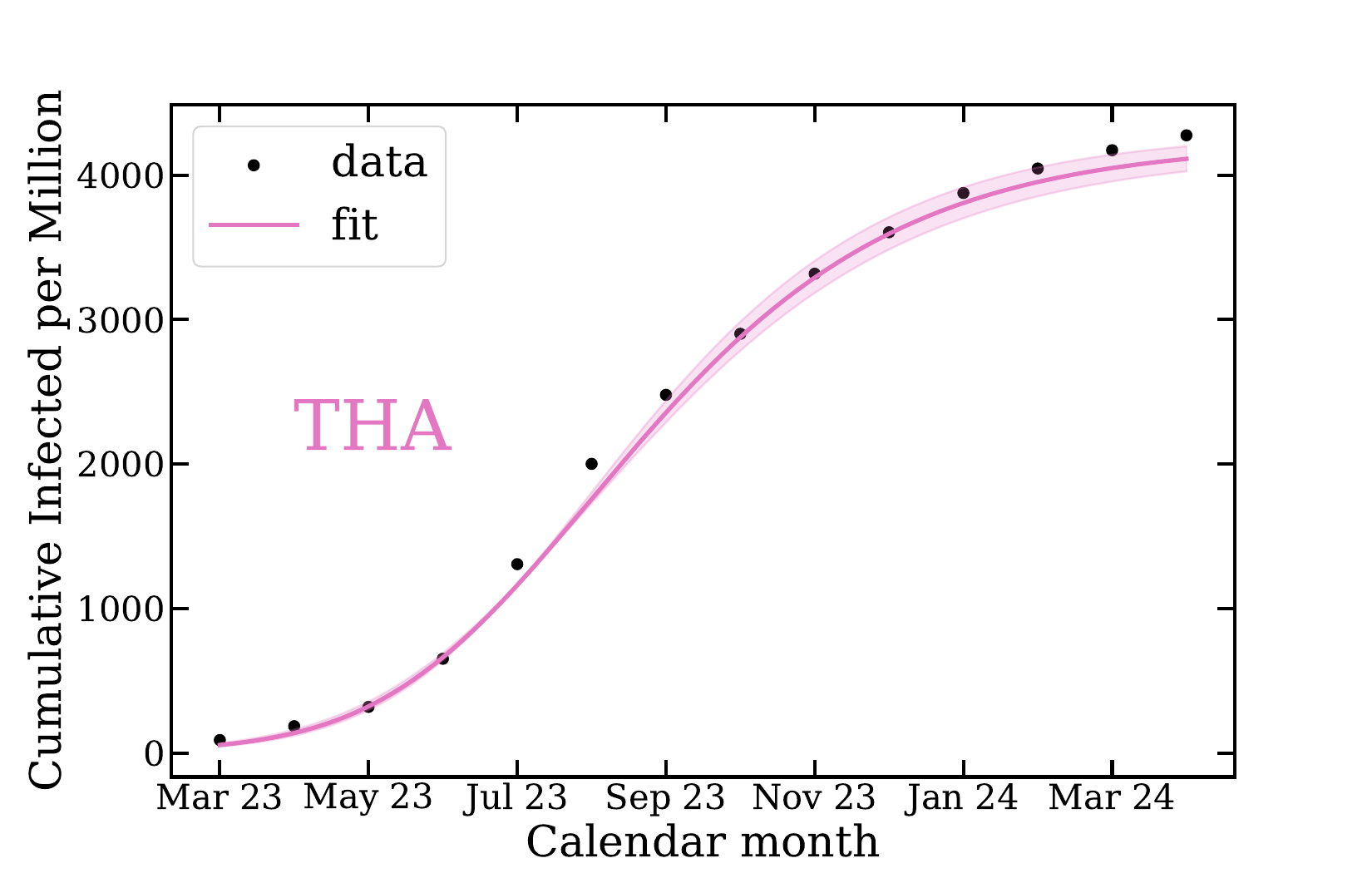}
		 \caption{THA}
		 \label{fig:THAFIT}
	      \end{subfigure}
    \caption{Plots of the official data of cumulative infected people per Million together with the eRG fit for the last wave.}
    \label{fig:cumulative_last_wave_asia}
\end{figure}

\FloatBarrier
The results of the eRG fit corresponding to the last wave are presented in Figure \ref{fig:cumulative_last_wave_asia}.

\FloatBarrier

\FloatBarrier

We now move on to provide the results of the eRG to both Nepal and Thailand. 

\paragraph{Nepal}
For Nepal, the fitted parameters obtained from the analysis of the most recent Dengue wave, that begins in March 2023, are:
\begin{equation}
  a =  7.48 \pm 0.01,\qquad b = 21.39 \pm 5.01 \ , \qquad \gamma = 1.02 \pm 0.05\ . 
\end{equation}

\paragraph{Thailand}
In Thailand, the most recent Dengue wave begins in March 2023. Figure \ref{fig:cumulative_ondate} illustrates the progression of infected cases over time, along with a best-fit curve that aligns closely with the data. The fitted parameters obtained are
\begin{equation}
  a =  8.35 \pm 0.02,\qquad b = 1.08 \pm 0.11 \ , \qquad \gamma = 0.44 \pm 0.02\ . 
\end{equation}

\begin{figure}[h!]
      \centering
	     \begin{subfigure}{0.45\linewidth}
		 \includegraphics[width=\linewidth]{ 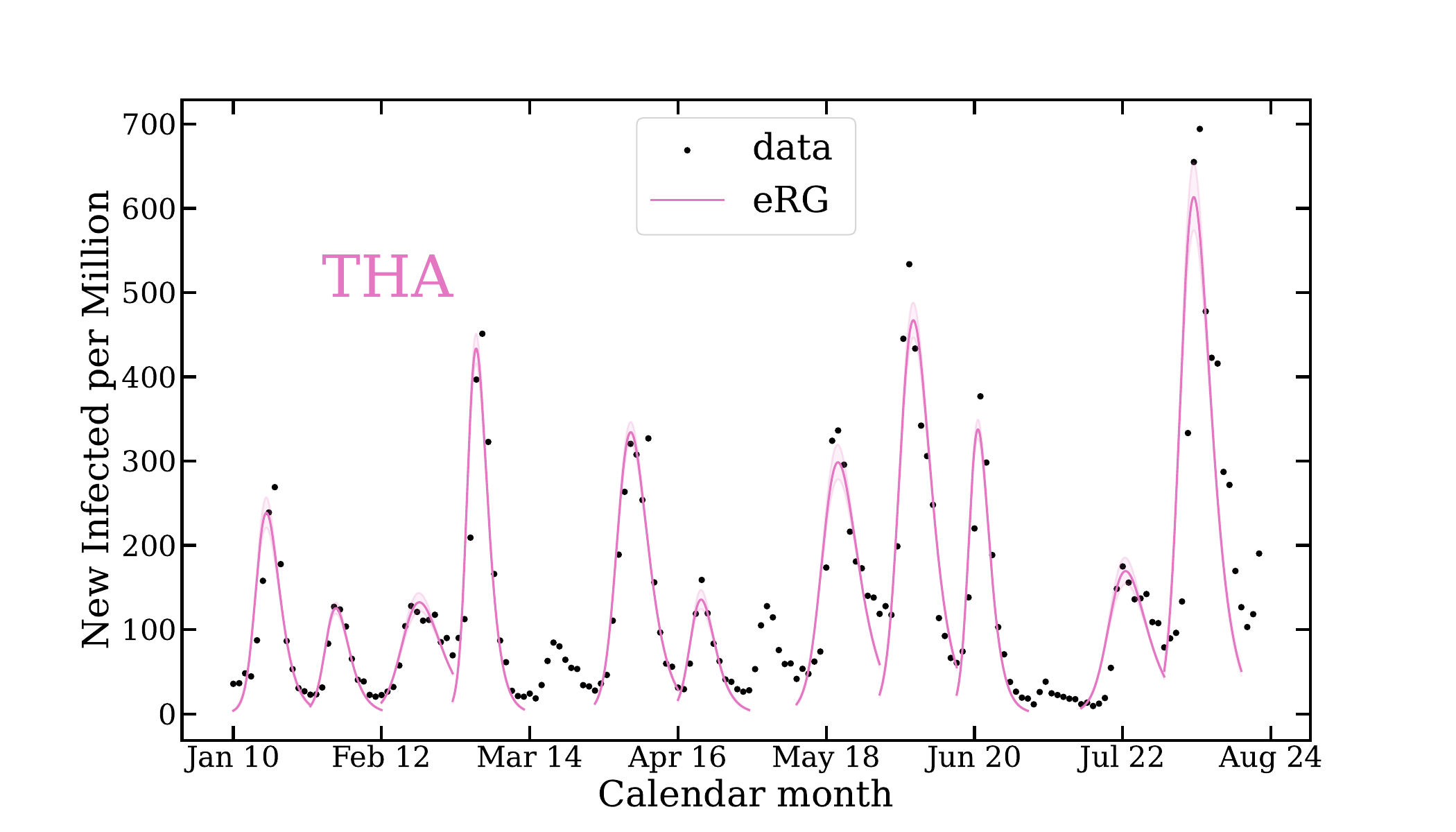}
		 \caption{THA}
		 \label{fig:THAni}
	      \end{subfigure}
       \begin{subfigure}{0.45\linewidth}
		 \includegraphics[width=\linewidth]{ 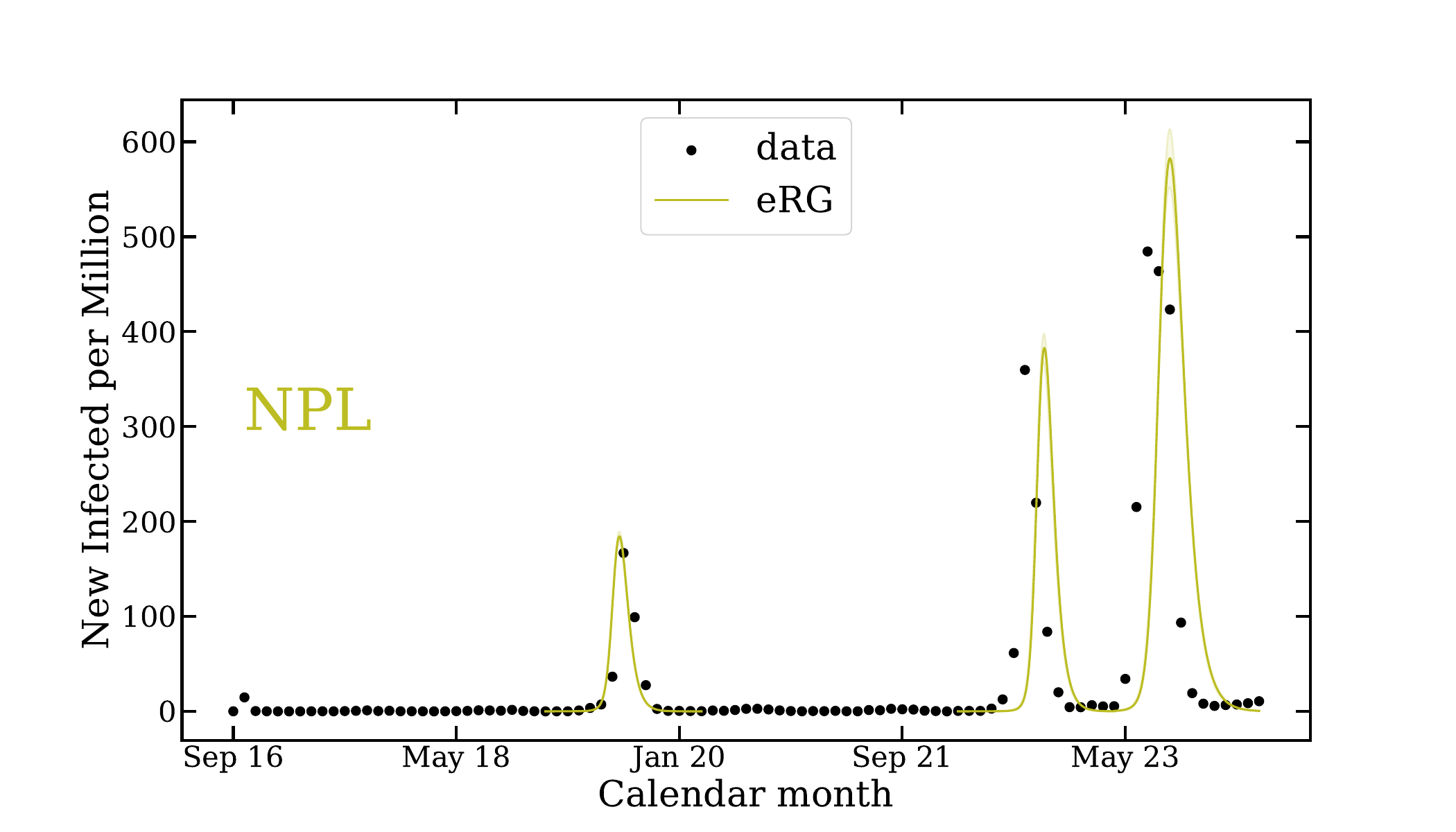}
		 \caption{NPL}
		 \label{fig:NPLni}
	      \end{subfigure}
	
    \caption{Plots of the official data of cumulative infected people per Million together with the eRG fit.}
    \label{fig:new_inf_wave_asia}
\end{figure}

In Figure \ref{fig:new_inf_wave_asia} we show the official number of new infected individuals over time in the two countries and the fitted results of the eRG description.

\FloatBarrier
\FloatBarrier
In Table \ref{tab:parameters_last_wave} we summarize the fitted parameters for all the investigated countries along with the corresponding one-sigma errors.
\FloatBarrier
\begin{table}[h!]
    \centering
    \begin{tabular}{|l|l|c|c|c|}
\hline \rowcolor{lightgray} \multicolumn{5}{|c|}{ \textbf{Recent Wave Parameters}} \\
\hline \rowcolor{lightgray} \textbf{Division} & \textbf{Code} & $a$ & $b$ & $\gamma$ \\
\hline Argentina & ARG & $9.47 \pm 0.06 $ & $9.92\pm 6.60$ & $0.94 \pm 0.15$ \\
\hline Bolivia & BOL  & $8.53 \pm 0.05 $ & $1.10\pm 0.15$ & $0.44 \pm 0.04$ \\
\hline Brazil & BRA  & $10.78 \pm 0.03 $ & $2.18\pm 0.33$ & $0.69 \pm 0.04$ \\
\hline Dominican Republic & DOM  & $8.03 \pm 0.02 $ & $2.29\pm5.50$ & $0.67 \pm 0.05$ \\
\hline Ecuador  & ECU & $8.01 \pm 0.08 $ & $0.85\pm0.50$ & $0.54 \pm 0.05$ \\
\hline Mexico  & MEX & $7.87 \pm 0.02 $ & $6.86\pm 1.53$ & $0.52 \pm 0.03$  \\
\hline
Nepal  & NPL & $7.48 \pm 0.01 $ & $21.39\pm 5.01$ & $1.02 \pm 0.05$  \\
\hline
Thailand  & THA & $8.35 \pm 0.02 $ & $1.08\pm 0.11$ & $0.44 \pm 0.02$  \\
\hline

\end{tabular}
    \caption{Parameters of the eRG model for the most recent Dengue wave in Latin American and Asian countries under study.}
    \label{tab:parameters_last_wave}
\end{table}

\FloatBarrier
As for the Latin American countries, we provide the fit result for all the Asian pandemic waves in Appendix \ref{tables}. It should be clear from the results that the eRG provides a good description of the pandemic waves for Asian countries as well.

\subsection{Global Warming impact on Dengue pandemic}
\label{sec:globalwarming}
It is widely recognized that climate change raises notable social concerns. Global warming significantly influences the spread of Dengue primarily by altering mosquitoes behavior and influencing their reproduction rates as temperatures increase. In particular, higher temperatures speed up the virus's growth within mosquitoes, increasing the risk of transmission.

Understanding how global warming influences Dengue outbreaks is crucial for creating effective prevention strategies. In this section, we explore the correlation between multi-wave Dengue patterns observed in Latin American and Asian countries and global warming trends. By employing temperature data from \url{https://www.wunderground.com/}, we confirm the expectation of a strong link between global warming and the worsening of Dengue outbreaks.

We therefore compare the temporal evolution of the eRG best-fit parameter $a$, measuring the total number of infected, with the average temperature of the most populated city in a given country. The results are summarized in Figure \ref{fig:a_and_temperature}, for all countries for which the temperature data were available. It is clear from the figures that the Dengue pandemic is highly impacted by the change in temperature and closely follows its variations, with the exception of Bolivia. However, for the latter one can still appreciate the overall trend of an increasing number of infected with the increase of temperature. We have further explored the relationship between the infection rate $\gamma$ and temperature for the same countries, but found that it is not significant. This could be due to the fact that the change in temperature does not significantly affect the intrinsic properties of the virus and the number of individuals nearby.

Overall, our study reinforces the expectation of the dangerous impact that  global warming has on the Dengue pandemic.

\FloatBarrier
\begin{figure}[h!]
      \centering
	     \begin{subfigure}{0.45\linewidth}
		 \includegraphics[width=\linewidth]{ 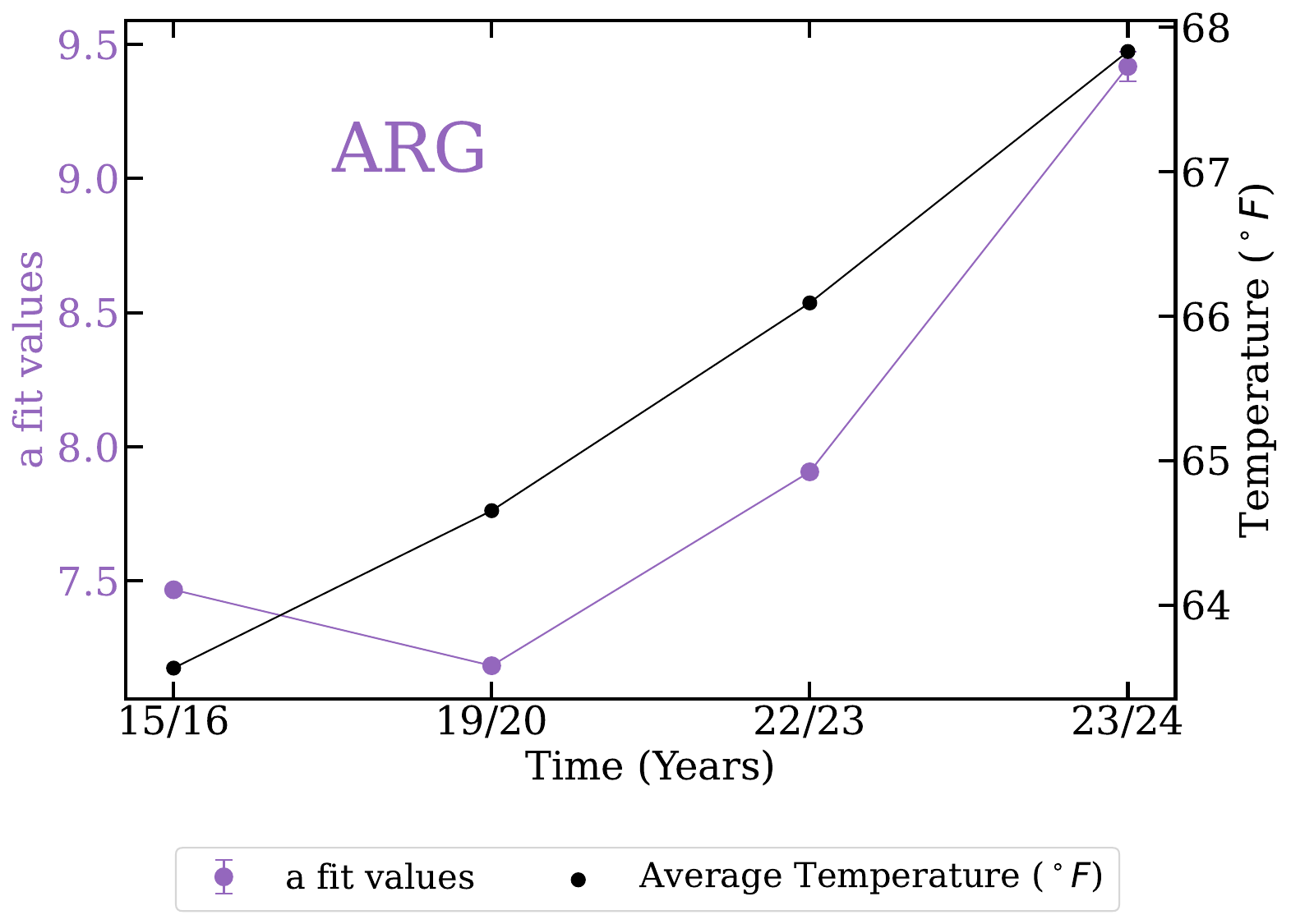}
		 \caption{ARG}
		 \label{fig:ARGtemp}
	      \end{subfigure}
	     \begin{subfigure}{0.45\linewidth}
		 \includegraphics[width=\linewidth]{ 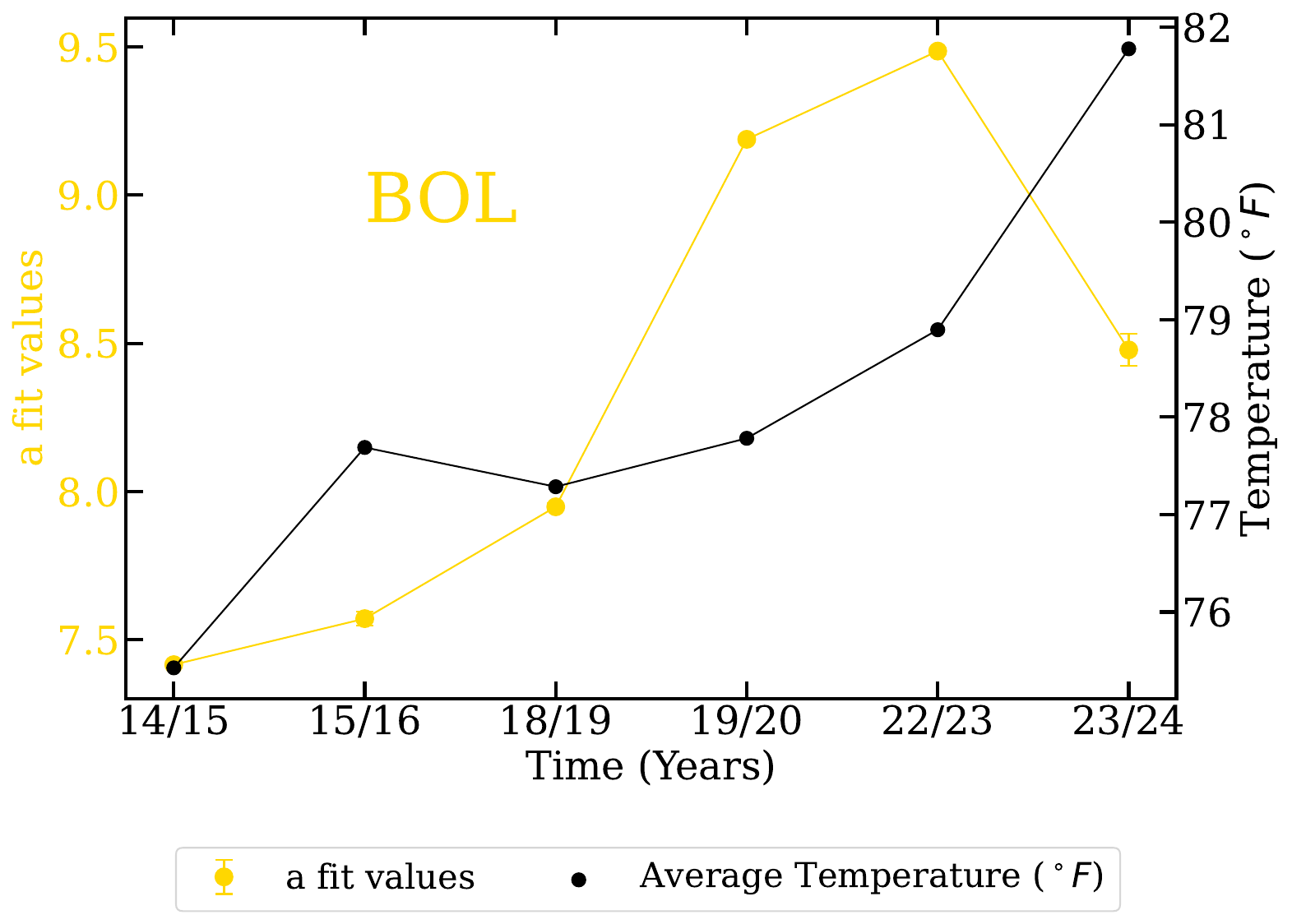}
		 \caption{BOL}
		 \label{fig:BOLtemp}
	      \end{subfigure}
      % \label{fig:plotsPartVax}
               \vfill
        \medskip
         \begin{subfigure}{0.45\linewidth}
    		  \includegraphics[width=\linewidth]{ 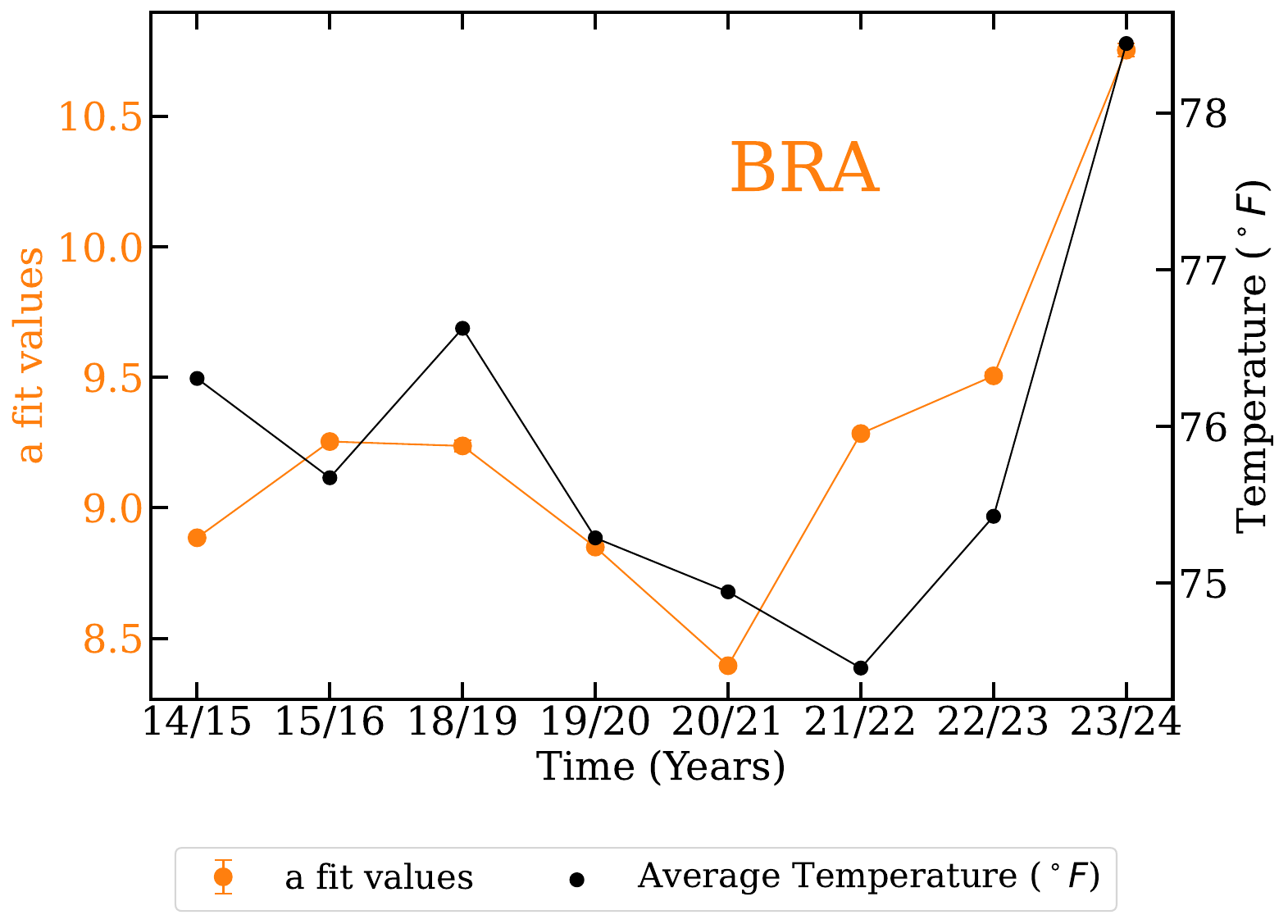}
    		  \caption{BRA}
    		  \label{fig:BRAtemp}
    	       \end{subfigure}
            \begin{subfigure}{0.45\linewidth}
    		  \includegraphics[width=\linewidth]{ 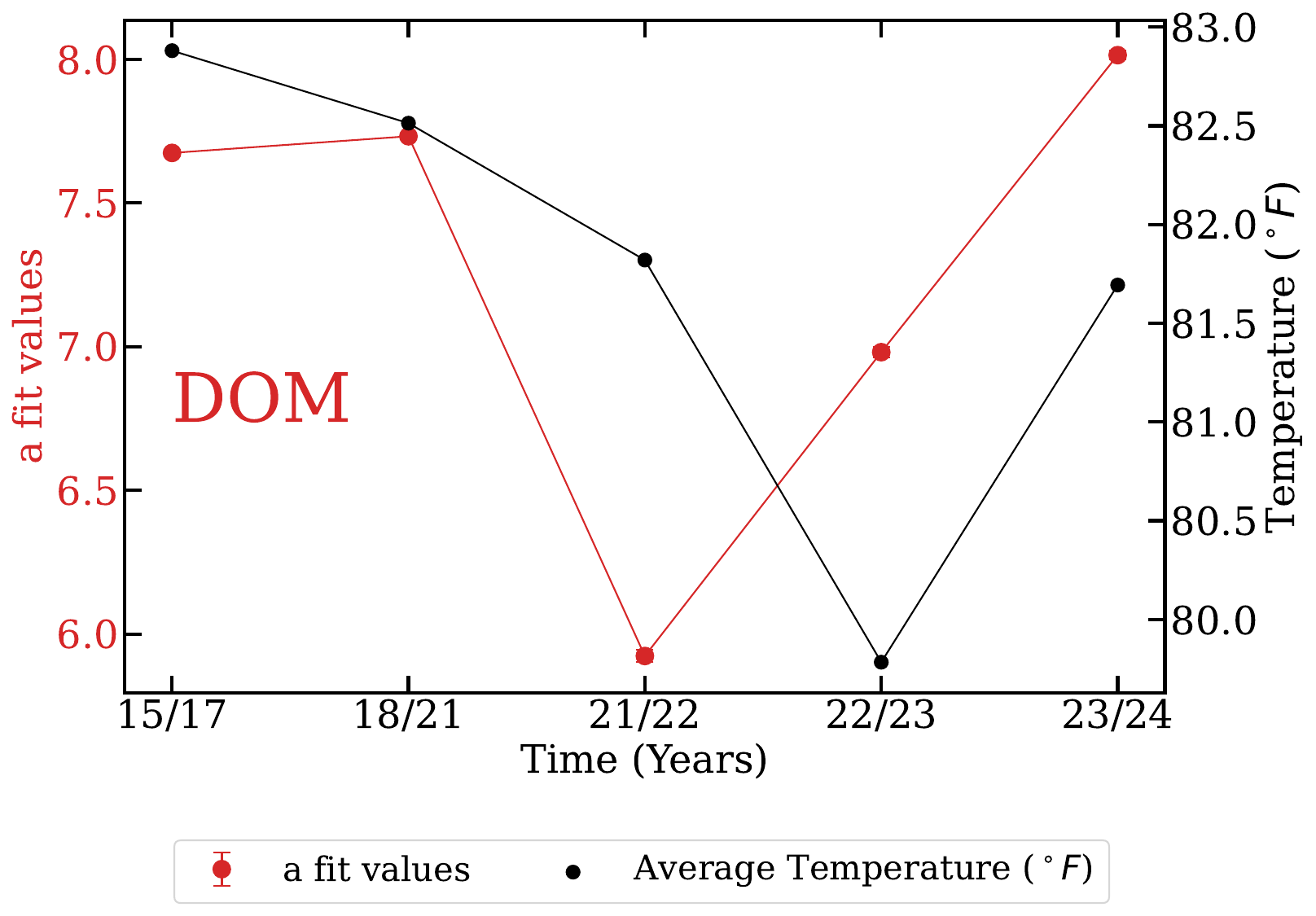}
    		  \caption{DOM}
    		  \label{fig:DOMtemp}
    	       \end{subfigure}
            \vfill
        \medskip
        \begin{subfigure}{0.45\linewidth}
    		  \includegraphics[width=\linewidth]{ 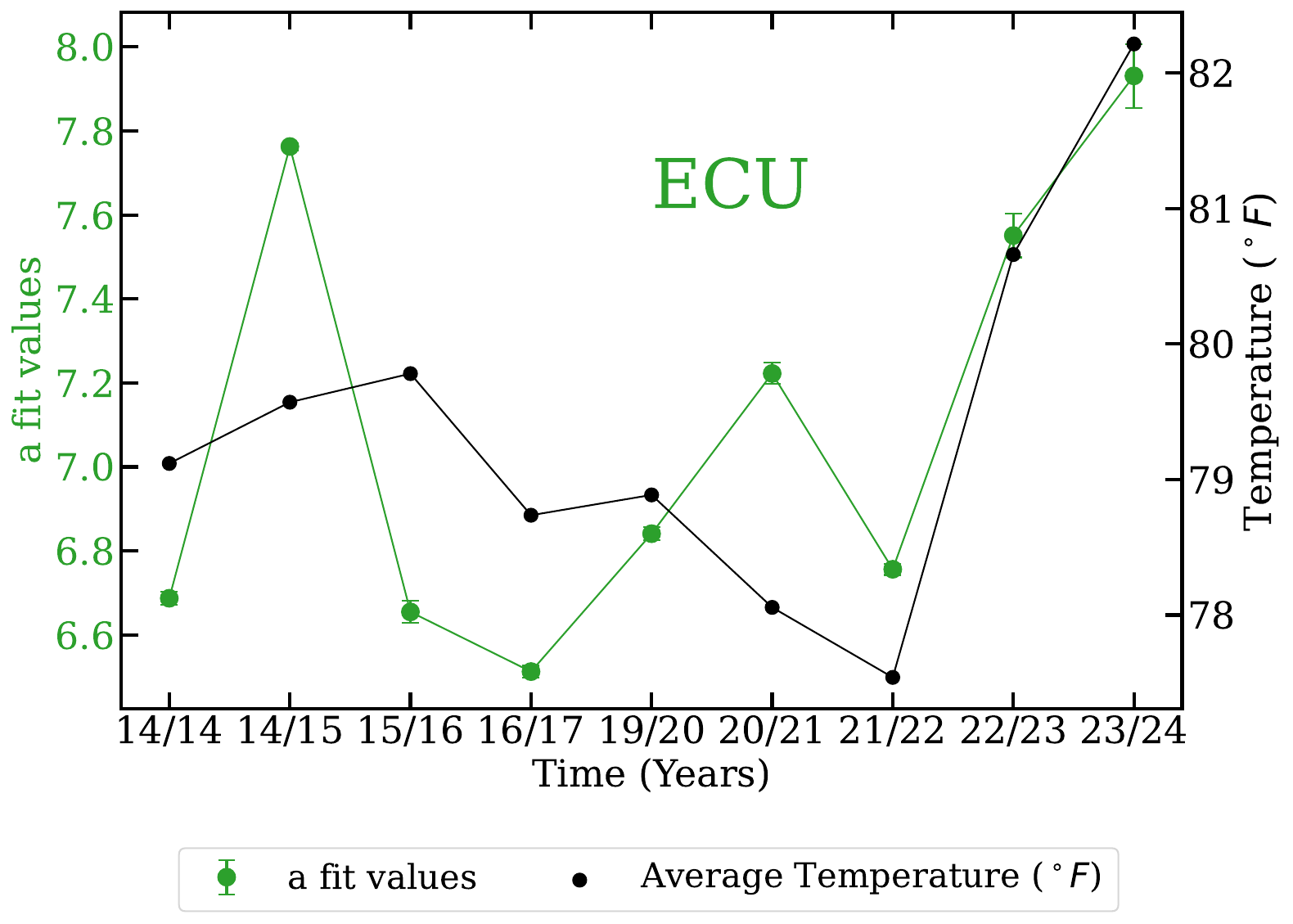}
    		  \caption{ECU}
    		  \label{fig:ECUtemp}
    	       \end{subfigure}
            \begin{subfigure}{0.45\linewidth}
    		  \includegraphics[width=\linewidth]{ 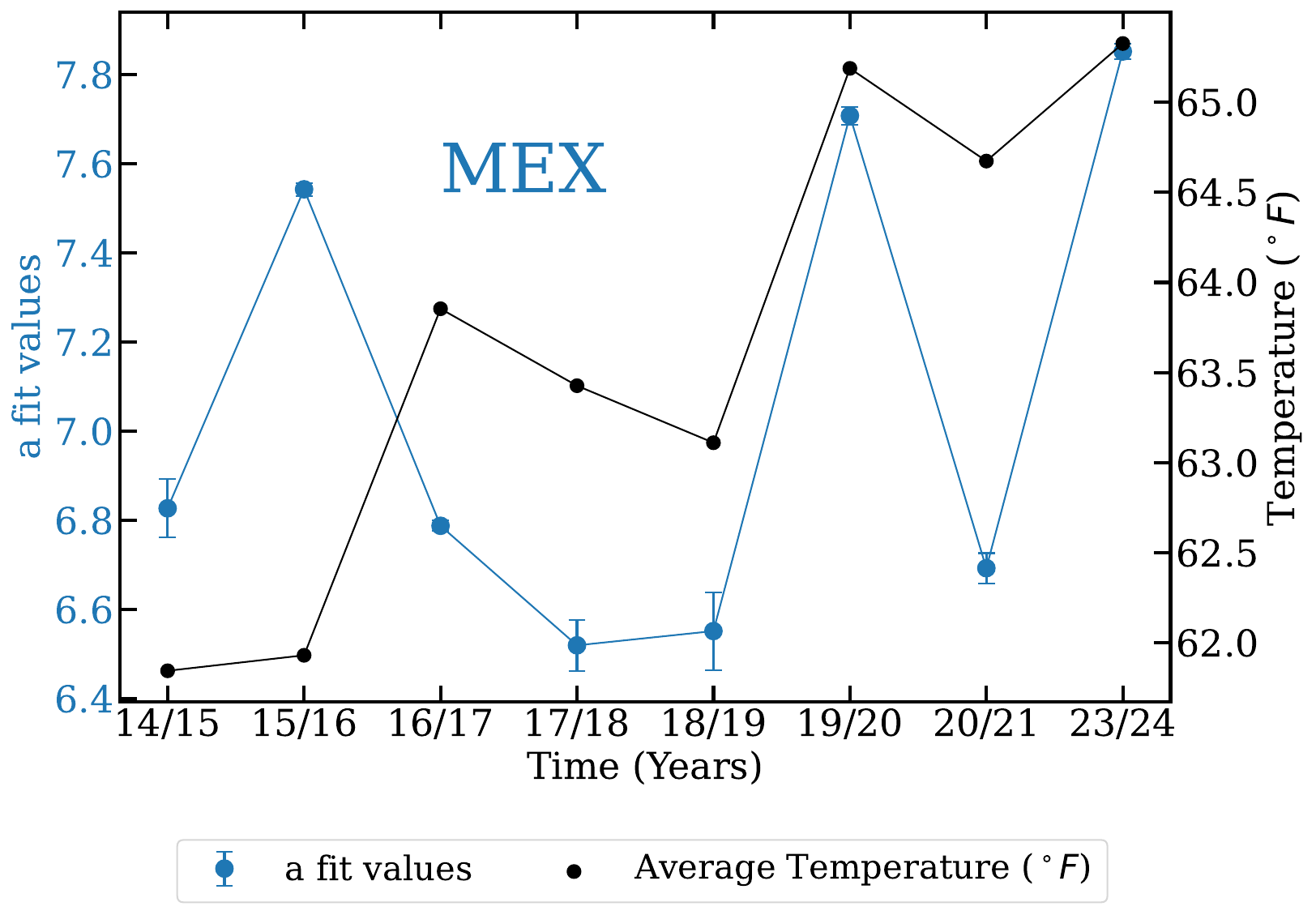}
    		  \caption{MEX}
    		  \label{fig:MEXtemp}
    	       \end{subfigure}
            \caption*{}
\end{figure}
\begin{figure}[h!]\ContinuedFloat
    \centering
         \begin{subfigure}{0.45\linewidth}
    		  \includegraphics[width=\linewidth]{ 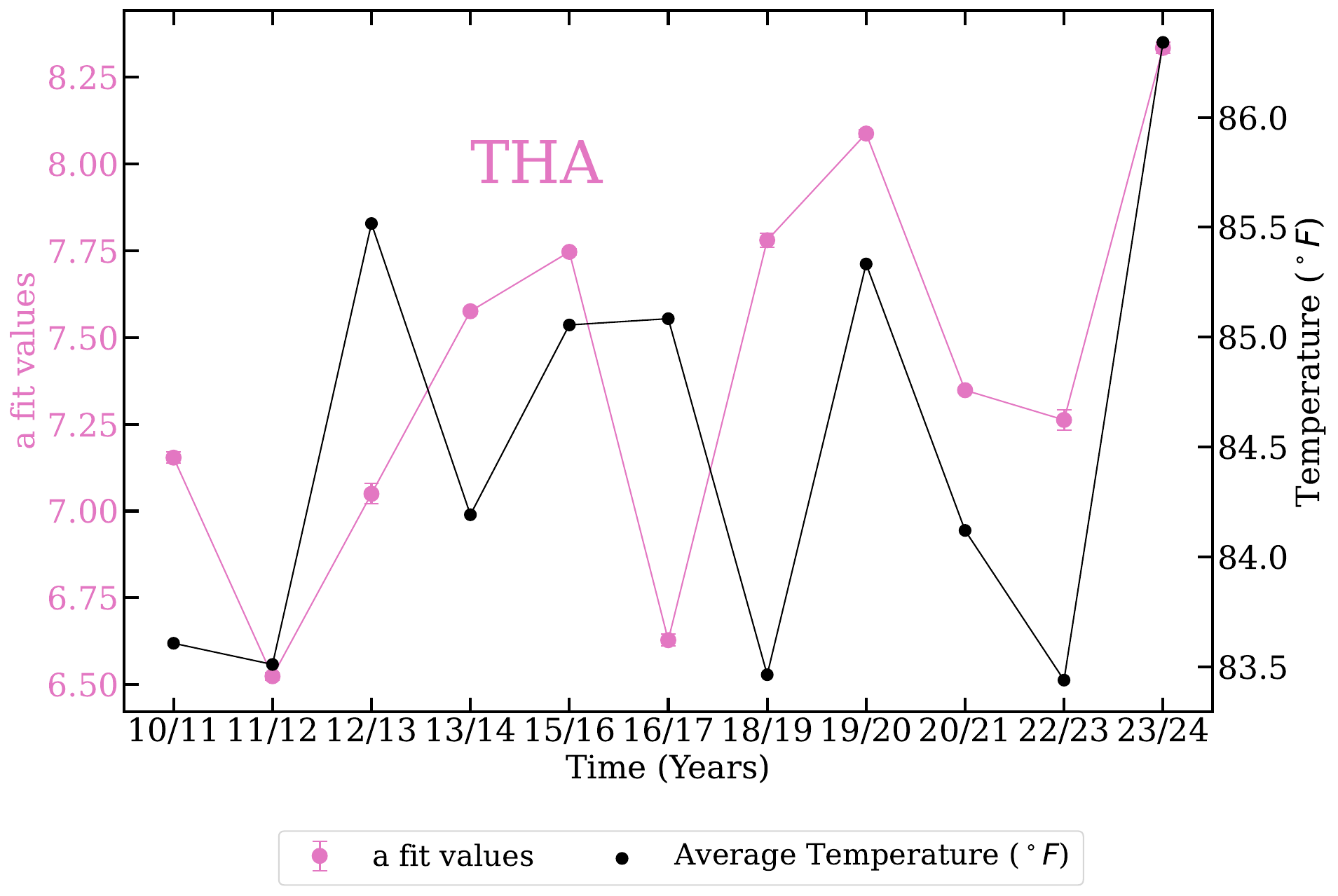}
    		  \caption{THA}
    		  \label{fig:THAtemp}
    	       \end{subfigure}
    \caption{Plots of the trend of both the best-fit $a$-parameter and temperature over the various epidemic waves. The temperature is averaged over the entire latter period.}
    \label{fig:a_and_temperature}
\end{figure}

\FloatBarrier

\section{Dengue in Italy: the case study of Fano }
\label{sec:fano}
In August 2024, an alarming outbreak of Dengue fever was officially reported in Fano, a municipality in the Marche region of Italy. With over 200 cases, including both confirmed and suspected infections, the situation has escalated to a significant public health concern. The emergence of this outbreak has raised fears for individual health of the whole community, prompting urgent action from health authorities and local governments.
 
%The unfolding situation in Fano serves as a critical reminder of the potential public health threats posed by the spread of Dengue disease, positioning it as an important case study for public health preparedness across Italy.

\begin{figure}[h!]
      \centering
	     \begin{subfigure}{0.45\linewidth}
		 \includegraphics[width=\linewidth]{ 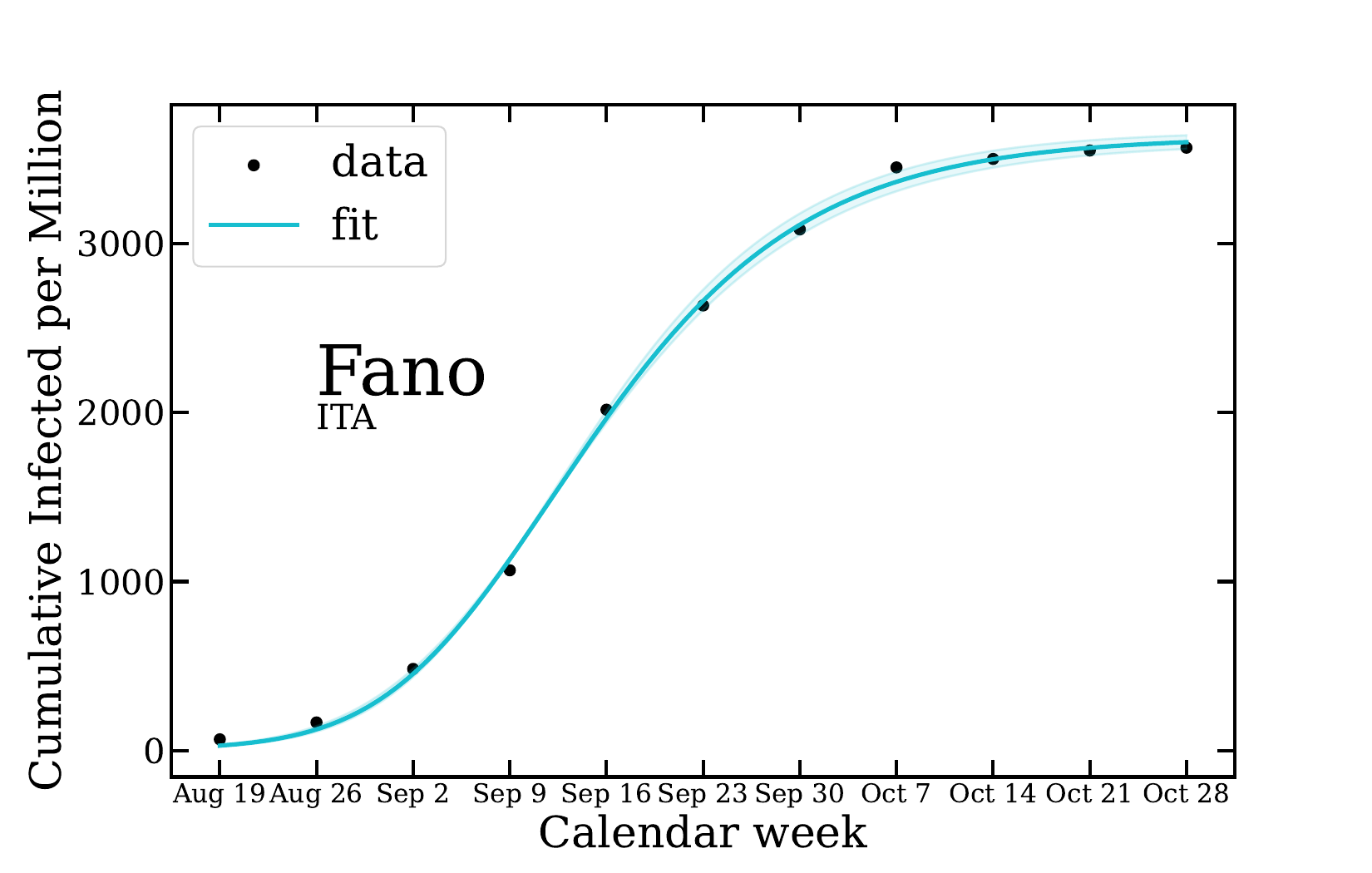}
		 \caption{Fit to the cumulative number of infected individuals per Million}
		 \label{fig:FANOfit}
	      \end{subfigure}
       \begin{subfigure}{0.45\linewidth}
		 \includegraphics[width=\linewidth]{ 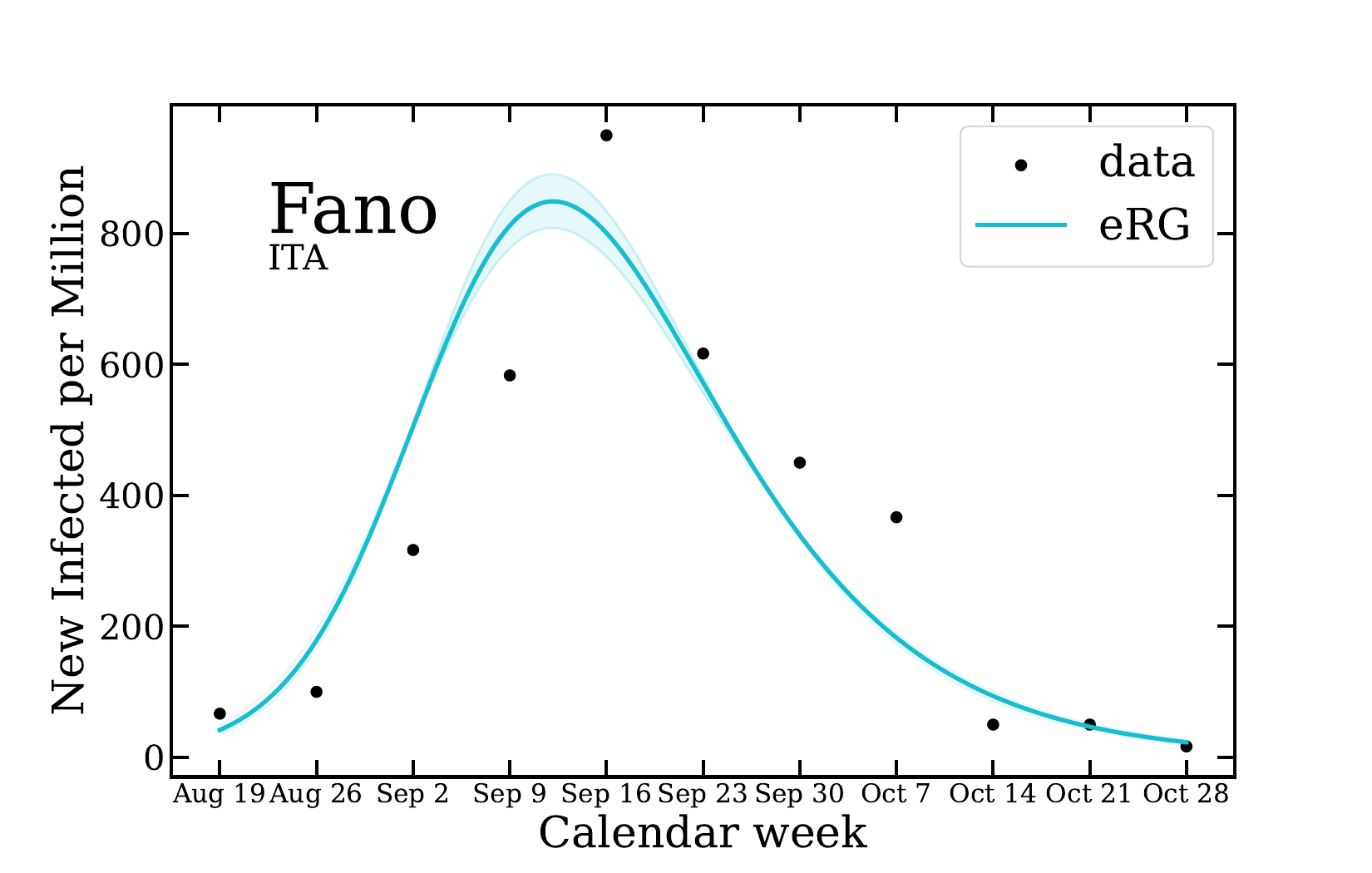}
		 \caption{eRG description of the new infected cases per Million.}
		 \label{fig:FANOnew}
	      \end{subfigure}
	
    \caption{Plots of the fit on the cumulative number of infected individuals per Million (Figure \ref{fig:FANOfit}) and the corresponding eRG curve for the new infected individuals per Million (Figure \ref{fig:FANOnew}) over time.}
    \label{fig:FANO}
\end{figure}

We now apply the eRG to describe the Dengue outbreak in Fano using the official data from the weekly report updated to October 31, 2024, available at \url{https://www.regione.marche.it/}. We perform the fit to the data normalized by the population of the town in millions. The results are shown in Figure \ref{fig:FANO}.

\FloatBarrier
\FloatBarrier
The best-fit parameters are
\begin{equation}
  a =  8.20 \pm 0.01,\qquad b =  2.92 \pm 0.36 \ , \qquad \gamma = 0.72 \pm 0.03\ . 
\end{equation}
 Since in Subsection \ref{sec:globalwarming} we identified a correlation between temperature and the parameters of our model we  extrapolate the parameter values describing potential Dengue outbreaks for the upcoming year in Fano. \\ In practice, we perform a linear fit on the reported monthly temperatures in Fano taken from \url{https://www.wunderground.com/} for the last three years and estimate the annual temperatures $T_{24},T_{25}$ for the years 2024 and 2025. The results in Subsection \ref{sec:globalwarming} suggest that it is possible to provide a naive estimate for the eRG parameter $a$ assuming the following linear relation with the temperature:
 \begin{equation}
     a_{\text{est}} = a \left(1 + \rho \frac{T_{25} - T_{24}} {T_{24}}\right) \ .
 \end{equation}
We estimate the parameter $\rho$ by averaging the Pearson correlation coefficient \cite{pearson1895note} obtained from the last three years of data for Argentina, Bolivia, Brazil, Dominican Republic, Ecuador, Mexico, and Thailand. This yields $\rho = 0.58 \pm 0.28$ that we take to predict the estimated $a$ parameter for the 2025 Fano pandemic wave:  
\begin{equation}
    a_{\text{est}} = 8.26\pm 0.14\ ,
\end{equation}
 which in turn translates into the following expected non-normalized cumulative number of infected individuals  
\begin{equation}
    \mathcal{I}_{\text{est}} = 232 \pm 32 \ .
\end{equation}  

%%%%%%%%%%%%%%%%%%%%%%%%%%%%%%%%%%%%%% to be written %%%%%%%%%%%%%%%%%%%%%%%%
 \section{Conclusions} \label{sec:conclusions}
 
In this work we investigated the temporal evolution of the Dengue pandemic from Latin American countries to Asian ones. To acquire solid estimates from the  data we employ  the  \emph{epidemiological Renormalization Group} (eRG) effective framework. We discovered that the eRG provides a simple but excellent description of the multiple waves Dengue dynamics across the globe. We further observe, in the Appendix, that the highest number of infected per million inhabitants is the highest in Brazil and Argentina. Additionally, the infection rate $\gamma$ is highest in Nepal with the next-to-highest in Argentina. Intriguingly, we showed that for the last few years $a$ tends to grow while typically $\gamma$ decreases. Via the acquired information we then noticed the correlation between the total number of infected individuals and the changes in the local temperature.

Because of the recent alarm triggered by the Dengue outbreak in Fano, Italy, we  employed the eRG to describe the current situation and then estimated the  impact of global warming to forecast the impact of Dengue for 2025.

 Our model can be employed to describe the spreading of other Arboviruses disease worldwide. Furthermore one can also extend it to take into account human behaviour along the lines of \cite{buonomo2024informationindexaugmentederg} and be used to investigate the gender impact.

\paragraph*{Acknowledgements}  

This research was supported by EU funding within the NextGenerationEU---MUR PNRR Extended Partnership initiative on Emerging Infectious Diseases (Project no. PE00000007, INF-ACT).
 The work of F.S. is
partially supported by the Carlsberg Foundation, semper ardens grant CF22-0922.

\newpage
\appendix

\section{Best-fit results for different countries and Dengue waves}
\label{tables}
In this Appendix we report the best-fit parameters for each epidemic wave and for each country. In the tables we report the month and the year of the start and end of each pandemic wave. 

%%%%%%%%% ARG %%%%%%%%%%%%%%%%%%%%5
\FloatBarrier
\begin{table}[h!]
    \centering
    \begin{tabular}{|l|c|c|}
\hline \rowcolor{argentina!70} \multicolumn{3}{|c|}{ \textbf{Argentina (ARG)}} \\
\hline \rowcolor{argentina!70} \textbf{Wave Period} & $a$  & $\gamma$ \\
\hline  $09/2015 - 11/2016$ & $7.47 \pm 0.00 $  & $1.10 \pm 0.02$ \\
\hline  $10/2019 - 10/2020$  & $7.19 \pm 0.01 $  & $1.49 \pm 0.10$ \\
\hline
$10/2022 - 10/2023$   & $7.91\pm 0.00$ & $1.41 \pm 0.02$ \\
\hline
$10/2023 - 06/2024$  & $9.47 \pm 0.06 $  & $0.94 \pm 0.15$ \\
\hline
\end{tabular}
    \caption{Parameters of the eRG model for waves in Argentina}
    \label{tab:parameters_ARG}
\end{table}
\FloatBarrier

%%%%%%%%%%%%%%%%%%%%%%%%%%%%% BOL %%%%%%%%%%%%%%%%%%%%

\FloatBarrier
\begin{table}[h!]
    \centering
    \begin{tabular}{|l|c|c|}
\hline \rowcolor{bolivia!70} \multicolumn{3}{|c|}{ \textbf{Bolivia (BOL)}} \\
\hline \rowcolor{bolivia!70} \textbf{Wave Period} & $a$  & $\gamma$ \\
\hline  $01/2014 - 09/2014$ & $7.43 \pm 0.01 $  & $1.03 \pm 0.06$ \\
\hline  $09/2014 - 08/2015$  & $7.60 \pm 0.02 $  & $1.06 \pm 0.11$ \\
\hline
$08/2015 - 09/2016$   & $7.96\pm 0.01$ & $0.83 \pm 0.06$ \\
\hline
$08/2019 - 08/2020$   & $9.20\pm 0.01$ & $1.09 \pm 0.04$ \\
\hline
$09/2022 - 09/2023$  & $9.49 \pm 0.01 $  & $0.97 \pm 0.04$ \\
\hline
$09/2023 - 06/2024$  & $8.53 \pm 0.05 $  & $0.44 \pm 0.04$ \\
\hline
\end{tabular}
    \caption{Parameters of the eRG model for waves in Bolivia.}
    \label{tab:parameters_BOL}
\end{table}
\FloatBarrier

%%%%%%%%%%%%%%%%%%%%%%%%%%%%% BRA %%%%%%%%%%%%%%%%%%%%%%%%%5

\FloatBarrier
\begin{table}[h!]
    \centering
    \begin{tabular}{|l|c|c|}
\hline \rowcolor{brazil!70} \multicolumn{3}{|c|}{ \textbf{Brazil (BRA)}} \\
\hline \rowcolor{brazil!70} \textbf{Wave Period} & $a$  & $\gamma$ \\
\hline  $10/2014 - 10/2015$ & $8.89 \pm 0.01 $  & $0.85 \pm 0.04$ \\
\hline  
$10/2015 - 10/2016$   & $9.26\pm 0.01$ & $0.86 \pm 0.03$ \\
\hline
$10/2018 - 10/2019$  & $9.26 \pm 0.02 $  & $0.69 \pm 0.06$ \\
\hline
$10/2019 - 10/2020$  & $8.86 \pm 0.01 $  & $0.62 \pm 0.01$ \\
\hline
$10/2020 - 10/2021$  & $8.40 \pm 0.01 $  & $0.54 \pm 0.02$ \\
\hline
$10/2021 - 10/2022$  & $9.30 \pm 0.01 $  & $0.68 \pm 0.04$ \\
\hline
$10/2022 - 10/2023$  & $9.52 \pm 0.01 $  & $0.73 \pm 0.04$ \\
\hline
$10/2023 - 06/2024$  & $10.78 \pm 0.03 $  & $0.69 \pm 0.04$ \\
\hline
\end{tabular}
    \caption{Parameters of the eRG model for waves in Brazil.}
    \label{tab:parameters_BRA}
\end{table}
\FloatBarrier

%%%%%%%%%%%%%%%%%%%%%%5 DOM %%%%%%%%%%%%%%%%%%%55

\FloatBarrier
\begin{table}[h!]
    \centering
    \begin{tabular}{|l|c|c|}
\hline \rowcolor{repdominican!70} \multicolumn{3}{|c|}{ \textbf{Dominican Republic (DOM)}} \\
\hline \rowcolor{repdominican!70} \textbf{Wave Period} & $a$  & $\gamma$ \\
\hline  $01/2015 - 11/2017$ & $7.68 \pm 0.01 $  & $0.35 \pm 0.01$ \\
\hline  $07/2018 - 05/2021$  & $7.74 \pm 0.01 $  & $0.38 \pm 0.01$ \\
\hline
$05/2021 - 02/2022$   & $5.95\pm 0.02$ & $0.54 \pm 0.02$ \\
\hline
$02/2022 - 05/2023$  & $7.00 \pm 0.02 $  & $0.42 \pm 0.02$ \\
\hline
$05/2023 - 06/2024$  & $8.03 \pm 0.02 $  & $0.67 \pm 0.05$ \\
\hline
\end{tabular}
    \caption{Parameters of the eRG model for waves in Dominican Republic.}
    \label{tab:parameters_DOM}
\end{table}
\FloatBarrier

%%%%%%%%%%%%%%%%%%%%%%% ECU %%%%%%%%%%%%%%%%%%%%%%%%%

\FloatBarrier
\begin{table}[h!]
    \centering
    \begin{tabular}{|l|c|c|}
\hline \rowcolor{ecuador!70} \multicolumn{3}{|c|}{ \textbf{Ecuador (ECU)}} \\
\hline \rowcolor{ecuador!70} \textbf{Wave Period} & $a$  & $\gamma$ \\
\hline  $01/2014 - 11/2014$ & $6.70 \pm 0.02 $  & $0.56 \pm 0.03$ \\
\hline  
$11/2014 - 11/2015$   & $7.77\pm 0.01$ & $0.86 \pm 0.04$ \\
\hline
$11/2015 - 10/2016$  & $6.68 \pm 0.03 $  & $0.46 \pm 0.03$ \\
\hline
$10/2016 - 10/2017$  & $6.53 \pm 0.01 $  & $0.57 \pm 0.03$ \\
\hline
$09/2019 - 11/2020$  & $6.86 \pm 0.02 $  & $0.46 \pm 0.02$ \\
\hline
$11/2020 - 12/2021$  & $7.25 \pm 0.01 $  & $0.36 \pm 0.03$ \\
\hline
$12/2021 - 12/2022$  & $6.77 \pm 0.01 $  & $0.52 \pm 0.03$ \\
\hline
$12/2022 - 12/2023$  & $7.60 \pm 0.05 $  & $0.26 \pm 0.02$ \\
\hline
$12/2023 - 06/2024$  & $8.01 \pm 0.08 $  & $0.53 \pm 0.05$ \\
\hline
\end{tabular}
    \caption{Parameters of the eRG model for waves in Ecuador.}
    \label{tab:parameters_ECU}
\end{table}
\FloatBarrier

%%%%%%%%%%%%%%%%%%%%%%%%%5 MEX %%%%%%%%%%%%%%%%%%%

\FloatBarrier
\begin{table}[h!]
    \centering
    \begin{tabular}{|l|c|c|}
\hline \rowcolor{mexico!70} \multicolumn{3}{|c|}{ \textbf{Mexico (MEX)}} \\
\hline \rowcolor{mexico!70} \textbf{Wave Period} & $a$  & $\gamma$ \\
\hline  $04/2014 - 01/2015$ & $6.89 \pm 0.07 $  & $0.49 \pm 0.06$ \\
\hline  
$01/2015 - 04/2016$   & $7.56\pm 0.01$ & $0.46 \pm 0.02$ \\
\hline
$04/2016 - 04/2017$  & $6.80 \pm 0.01 $  & $0.66 \pm 0.03$ \\
\hline
$04/2017 - 01/2018$  & $6.58 \pm 0.06 $  & $0.60 \pm 0.08$ \\
\hline
$01/2018 - 01/2019$  & $6.64 \pm 0.09 $  & $0.44 \pm 0.07$ \\
\hline
$01/2019 - 04/2020$  & $7.73 \pm 0.02 $  & $0.69 \pm 0.06$ \\
\hline
$04/2020 - 02/2021$  & $6.73 \pm 0.03 $  & $0.61 \pm 0.06$ \\
\hline
$01/2023 - 04/2024$  & $7.87 \pm 0.02 $  & $0.52 \pm 0.03$ \\
\hline
\end{tabular}
    \caption{Parameters of the eRG model for waves in Mexico.}
    \label{tab:parameters_MEX}
\end{table}
\FloatBarrier

%%%%%%%%%%%%%%%%%%55 NPL %%%%%%%%%%%%%%%%%%%%%%%%%%%%%%%%

\FloatBarrier
\begin{table}[h!]
    \centering
    \begin{tabular}{|l|c|c|}
\hline \rowcolor{nepal!70} \multicolumn{3}{|c|}{ \textbf{Nepal (NPL)}} \\
\hline \rowcolor{nepal!70} \textbf{Wave Period} & $a$  & $\gamma$ \\
\hline  $02/2019 - 03/2020$ & $5.85 \pm 0.01 $  & $1.72 \pm 0.04$ \\
\hline  $01/2022 - 03/2023$  & $6.64 \pm 0.01 $  & $1.58 \pm 0.06$ \\
\hline
$03/2023 - 04/2024$   & $7.48\pm 0.15$ & $1.02 \pm 0.05$ \\
\hline
\end{tabular}
    \caption{Parameters of the eRG model for waves in Nepal.}
    \label{tab:parameters_NPL}
\end{table}
\FloatBarrier

%%%%%%%%%%%%%%%%%%%%%%%%%%%%% THA %%%%%%%%%%%%%%%%%%%%%%%%%%%%

\FloatBarrier
\begin{table}[h!]
    \centering
    \begin{tabular}{|l|c|c|}
\hline \rowcolor{thailand!70} \multicolumn{3}{|c|}{ \textbf{Thailand (THA)}} \\
\hline \rowcolor{thailand!70} \textbf{Wave Period} & $a$  & $\gamma$ \\
\hline  $01/2010 - 02/2011$ & $7.17 \pm 0.02 $  & $0.57 \pm 0.03$ \\
\hline  $02/2011 - 02/2012$  & $6.54 \pm 0.01 $  & $0.58 \pm 0.02$ \\
\hline
$02/2012 - 02/2013$   & $7.08\pm 0.03$ & $0.35 \pm 0.02$ \\
\hline
$02/2013 - 02/2014$  & $7.58 \pm 0.01 $  & $0.68 \pm 0.02$ \\
\hline
$02/2015 - 04/2016$ & $7.76 \pm 0.01 $  & $0.44 \pm 0.01$ \\
\hline  
$04/2016 - 04/2017$  & $6.65 \pm 0.02 $  & $0.56 \pm 0.04$ \\
\hline
$12/2017 - 02/2019$   & $7.80\pm 0.02$ & $0.38 \pm 0.02$ \\
\hline
$01/2019 - 03/2020$  & $8.10 \pm 0.01 $  & $0.44 \pm 0.02$ \\
\hline
$03/2020 - 04/2021$  & $7.35 \pm 0.01 $  & $0.67 \pm 0.02$ \\
\hline
$01/2022 - 03/2023$ & $7.29 \pm 0.03 $  & $0.36 \pm 0.02$ \\
\hline  
$03/2023 - 04/2024$  & $8.35 \pm 0.02 $  & $0.44 \pm 0.02$ \\
\hline
\end{tabular}
    \caption{Parameters of the eRG model for waves in Thailand.}
    \label{tab:parameters_THA}
\end{table}
\FloatBarrier

One finds the highest number of infected per million inhabitants  in Brazil and  Argentina. The highest infection rate $\gamma$ is found in Nepal, followed by Argentina and Bolivia. Interestingly we discover that for the last years that while $a$ tends to grow typically $\gamma$ decreases.

\printbibliography[heading=bibintoc]
\end{document}